%% file: Great3D.tex
\documentclass[journal,10pt]{IEEEtran}
\IEEEoverridecommandlockouts
\input{setting-ieee}

\graphicspath{{./figs/}{../}}

\newcommand{\shorttitle}{\textsc{Great3D}}

\definecolor{blue}{rgb}{0,0,0}
\definecolor{red}{rgb}{0,0,0}
\renewcommand{\revise}[1]{#1}

\renewcommand{\DIFdel}[1]{}
\hypersetup{colorlinks=true, citecolor=black, linkcolor=black, urlcolor=black}

\begin{document}

\title{

    Partitioning-free 3D-IC Floorplanning

}

\author{Shuo~Ren,~\hspace{6pt}Rongliang~Fu${^*}$,~\hspace{6pt}Libo~Shen,~\hspace{6pt}Zhen Zhuang,~\hspace{6pt}Leilei~Jin,~\hspace{6pt}\\
    Hao~Yu~\IEEEmembership{Senior Member,~IEEE},\hspace{6pt} Bei~Yu~\IEEEmembership{Senior Member,~IEEE},\hspace{6pt} Tsung-Yi~Ho~\IEEEmembership{Fellow,~IEEE}
    \thanks{A preliminary version of this paper is to be presented at the 2026 Asia and South Pacific Design Automation Conference (ASP-DAC~2026)~\cite{2026ASPDAC_Great3D}.}
    \IEEEcompsocitemizethanks{
        \IEEEcompsocthanksitem Shuo~Ren, Rongliang~Fu, Libo~Shen, Zhen Zhuang, Leilei~Jin, Bei Yu, and Tsung-Yi~Ho are with The Chinese University of Hong Kong.
        E-mail: \{sren, rlfu, lbshen24, zzhuang21, lljin, ~byu,~tyho\}@cse.cuhk.edu.hk.
        \IEEEcompsocthanksitem Hao~Yu is with the Southern University of Science and Technology, Shenzhen, China.
        E-mail: yuh3@sustech.edu.cn.
        \IEEEcompsocthanksitem  ${^*}$Corresponding author: Rongliang Fu.
    }
}

\maketitle
\thispagestyle{plain}
\pagestyle{plain}

\input{doc/0-abstract}

\begin{IEEEkeywords}
    3D IC, 3D Floorplanning, Partitioning-free, Semi-definite Programming
\end{IEEEkeywords}

\input{doc/1-intro}
\input{doc/2-prelim} %
\input{doc/3-Methodology}
\input{doc/4-experiment}

\input{doc/5-conclusion}

\bibliographystyle{IEEEtran}
\bibliography{ref/Top,ref/reference}

\input{doc/bio}

\end{document}

%% file: setting-ieee.tex
\usepackage{blkarray}                                      %
\usepackage{graphicx}                                      %
\usepackage{amsmath}
\usepackage{amssymb}
\usepackage{amsfonts}
\usepackage{amsthm}
\usepackage[mathcal]{eucal}
\usepackage{mathrsfs}
\usepackage{booktabs}
\usepackage{enumerate}
\usepackage{multirow}
\usepackage[subrefformat=parens,farskip=0pt,justification=centering]{subfig}
\usepackage{color}
\usepackage{cite}                                          %
\usepackage{comment}                                       %
\usepackage{soul}                                          %
\soulregister\cite7
\soulregister\ref7
\soulregister\pageref7
\usepackage{etoolbox}                                      %
\usepackage{url}
\usepackage{nth}                                           %
\usepackage{bm}                                            %
\usepackage{courier}
\usepackage{balance}
\usepackage{threeparttable}
\usepackage{xcolor,colortbl}
\usepackage{footnote}
\usepackage{listings}
\usepackage{setspace}                                      %
\usepackage[inline]{enumitem}
\usepackage{verbatim}
\usepackage[bookmarks=false]{hyperref}
\hypersetup{
    colorlinks = true,
    citecolor  = blue,
    linkcolor  = blue,
    urlcolor   = blue,
}
\usepackage{tikz}
\usetikzlibrary{patterns,snakes}
\usetikzlibrary{positioning,calc,fit,decorations.pathmorphing,shapes.geometric, shapes.gates.logic.US, calc}
\usetikzlibrary{arrows,arrows.meta,decorations.markings,shapes,shapes.arrows}
\usetikzlibrary{decorations,decorations.pathreplacing}
\usetikzlibrary{backgrounds}
\usepackage{filecontents}                                  %
\usepackage{pgfplots}
\usepackage{pgfplotstable}
\usepackage{scalefnt}
\pgfplotsset{compat=newest}
\usepackage{caption}
\usepackage{pifont}                                        %
\usepackage{cleveref}
\Crefformat{figure}{Fig.~#2#1#3}                           %
\Crefname{subfigure}{Fig.}{Figs.}
\Crefname{figure}{Fig.}{Figs.}
\Crefformat{table}{TABLE~#2#1#3}                           %
\usepackage[figuresright]{rotating}

\usepackage{algorithm}
\iftrue
\usepackage{algpseudocode}                                 %
\algrenewcommand\textproc{\texttt}
\makeatletter
\let\OldStatex\Statex
\renewcommand{\Statex}[1][3]{%
  \setlength\@tempdima{\algorithmicindent}%
  \OldStatex\hskip\dimexpr#1\@tempdima\relax
}
\makeatother
\else
\usepackage{algorithmic}
\fi

\definecolor{CUHKorange}{RGB}{244,106,18} %
\definecolor{CUHKblue}{RGB}{0,111,190}    %
\definecolor{CUHKgreen}{RGB}{0,127,128}   %
\definecolor{CUHKred}{RGB}{228,46,36}     %
\definecolor{CUHKyellow}{RGB}{198,148,34} %
\definecolor{CUHKdark}{RGB}{114,44,114}   %
\definecolor{CUHKmiddle}{RGB}{144,44,144} %
\definecolor{CUHKlight}{RGB}{167,44,167} 
\definecolor{CUHKpurple}{RGB}{117,15,109}
\definecolor{CUHKgold}{RGB}{221,163,0}
\definecolor{CUHKribbon}{RGB}{244,223,176}
\definecolor{CUHKblack}{RGB}{34,24,21}

\renewcommand{\bf}[1]{\textbf{#1}}

\newcommand{\mysection}[1]{\vspace{.06in}\noindent{\textbf{#1}}}

\usepackage{tcolorbox}
\tcbuselibrary{skins,breakable}
    {\endtcolorbox}
    {\endtcolorbox}

\usepackage[top=0.80in,bottom=0.88in,left=0.63in,right=0.63in]{geometry}
\crefname{mytheorem}{Theorem}{Theorems}
\crefname{mylemma}{Lemma}{Lemmas}
\crefname{myclaim}{Claim}{Claims}
\crefname{myproperty}{Property}{Properties}
\crefname{mycorollary}{Corollary}{Corollaries}

\RequirePackage[normalem]{ulem} %
\RequirePackage{color}\definecolor{RED}{rgb}{1,0,0}\definecolor{BLUE}{rgb}{0,0,1} %
\providecommand{\DIFdel}[1]{{\protect\color{red}\sout{#1}}}                       %

\newcommand{\revise}[1]{\DIFadd{#1}}

%% file: doc/0-abstract.tex
\begin{abstract}
  3D integration with fine-pitch hybrid bonding offers a promising path to
  alleviate interconnect bottlenecks in conventional two-dimensional (2D) ICs, yet
  efficient 3D floorplanning remains challenging due to the enlarged solution
  space and non-uniform inter-die communication latency. Existing methods either
  extend 2D representations into 3D, leading to combinatorial complexity, or adopt
  partitioning-first pipelines that fix block-to-die assignments early and hinder
  joint optimization of floorplan, die assignment, and vertical connectivity. In
  this work, we present \textsc{Great3D}, a partitioning-free 3D floorplanning
  framework that directly optimizes a native 3D floorplan. \textsc{Great3D}
  formulates a unified objective that couples interconnect cost with a
  cycles-per-instruction (CPI)-derived latency term to capture the system-level impact of face-to-face
  (F2F) bonding. Algorithmically, it combines an SDP-based 3D global embedding
  with a dynamic-programming refinement for die assignment, followed by 2D
  continuous refinement with practical design constraints.
  \textcolor{blue}{Experiments on the GSRC and ATPlace benchmark suites
  show that \textsc{Great3D} consistently achieves strong wirelength and CPI quality
  against state-of-the-art 3D floorplanners. On GSRC, it reduces total wirelength by
  up to about $70\%$ (and by $2.40$--$2.74\times$ on average) over competing 3D-native
  floorplanners, and its dynamic-programming die-assignment stage further improves CPI
  by $9.5$--$17.8\%$, while maintaining competitive runtime on instances of up to a few
  hundred blocks.}
\end{abstract}

%% file: doc/1-intro.tex
\section{Introduction}
\label{sec:intro}

\IEEEPARstart{W}{hile}
transistor scaling has largely sustained Moore’s law, interconnect scaling has
lagged behind, making routing delay a dominant performance bottleneck in modern
two-dimensional (2D) integrated circuits (ICs)~\cite{2019JSA-Cao}.
Three-dimensional (3D) integration alleviates this bottleneck by stacking device
layers vertically, transforming long intra-die wires into short inter-die
connections while increasing integration density~\cite{2025ASPDAC-towards3dic}.
Recent advances in fine-pitch hybrid bonding further enable TSV-free,
face-to-face (F2F) interconnects, which are now being adopted in
commercial 3D ICs~\cite{2024ECTC-hybrid-bonding,2024-cucu-hybridbonding,hybrid-cores-intel}.
These benefits, however, introduce new physical-design challenges: designers
must manage inter-die connectivity with non-uniform communication costs and
navigate a much larger 3D floorplan solution space~\cite{2025ASPDAC-towards3dic}.
Among all backend stages, floorplanning is particularly critical because it
simultaneously determines block locations and die assignments, which directly
impact both wirelength and communication latency in 3D stacks. Consequently,
efficient 3D floorplanning has become a key problem in modern IC design.

Unlike 2D floorplanning, which determines block locations within a single planar
layer, 3D floorplanning must simultaneously determine both in-die floorplan
locations and die assignments across multiple vertically stacked tiers. This
additional vertical dimension introduces fundamental new complexities, including
$z$-axis feasibility, cross-tier legality, and non-uniform communication
latency. In particular, inter-die links typically incur higher delay than
intra-die wires due to bonding interfaces and stack-level constraints, and thus
should be modeled explicitly during layout planning rather than treated as a
secondary effect. As a result, conventional 2D floorplanning techniques are not
directly applicable to 3D floorplanning.

\begin{figure}[t!]
      \centering
      \subfloat[Before optimization]{%
            \includegraphics[width=0.48\linewidth,trim={0 0 429pt 0},clip]{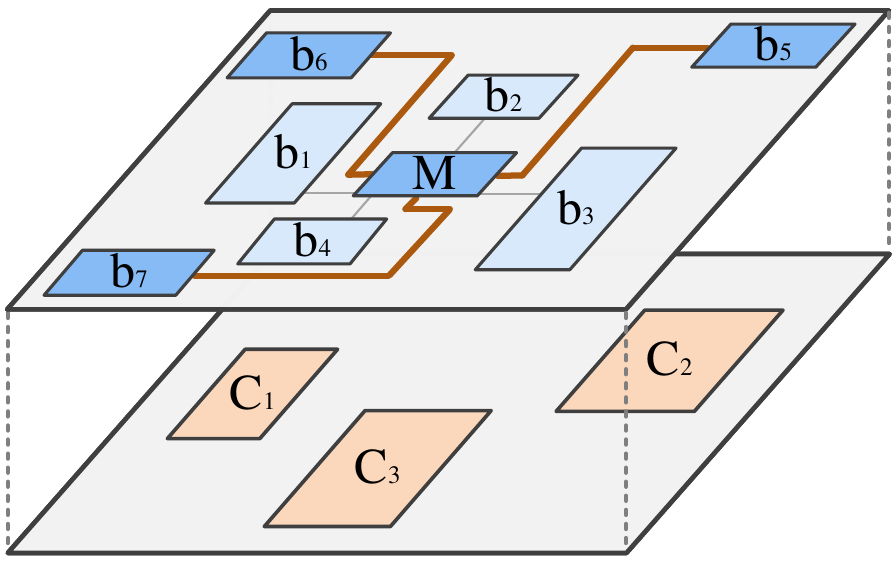}%
            \label{fig:intra2inter-before}%
      }
      \hfill
      \subfloat[After optimization (Ours)]{%
            \includegraphics[width=0.48\linewidth,trim={429.8pt 0 0 0},clip]{figs/intra2inter-v4.pdf}%
            \label{fig:intra2inter-after}%
      }
      \caption{Comparison of 3D floorplans before and after inter-die optimization. (a)~Baseline layout constrains highly connected modules to a single tier, limiting vertical integration. (b)~Our method redistributes blocks across tiers, enabling shorter inter-die connections and better area utilization under identical outline constraints.}
      \label{fig:intra2inter}
\end{figure}

Existing 3D floorplanning methods can be broadly grouped into two categories,
each making restrictive assumptions that limit scalability or solution quality.

The first category~\cite{2005ASPDAC-flpdatastructure-3dslicingtree,
      2011TVLSI-grouped-sequence-pair, 2010Integration-Partition-Sequence-pair,
      2013ASPDAC-monolithic-3d-floorplan,
      2024neurips-flexplanner} extends 2D representations into 3D by embedding die
assignment directly into the floorplan structure. For example, 3D slicing
trees~\cite{2005ASPDAC-flpdatastructure-3dslicingtree} enforce hierarchical
layer partitioning, while grouped sequence pairs~\cite{2011TVLSI-grouped-sequence-pair}
encode cross-layer relative ordering. FlexPlanner~\cite{2024neurips-flexplanner}
uses a graph-based representation learned from prior layout data, but its
performance can depend on the quality and generalizability of the training
set, which may limit its applicability to unseen architectures. Although these methods preserve a high degree of flexibility, they suffer from
exponential growth of the solution space and tightly coupled objectives: die
assignment and in-die floorplan locations are optimized together, forcing the
use of slow metaheuristics and making it difficult to reason explicitly about
inter-die latency.
In particular, the lack of an explicit, tractable latency-aware objective
and controllable feasibility handling makes it challenging to incorporate
system-level performance signals (e.g., cycles per instruction, CPI) and fixed-outline constraints in a
scalable optimization flow.

The second category~\cite{2024ASPDAC-coplace,
      2021TVLSI-thermal-aware-3dfloorplan-tsv, 2022TCAD-snap3d,
      panth2015placement-tcad, park2021pseudo} adopts a
partitioning-first paradigm. These approaches decompose the 3D task into
sequential 2D subproblems via pre-processing: a graph partitioner such as
Fiduccia–Mattheyses (FM)~\cite{1995ICCAD-FM_partition} or spectral partitioning
first assigns blocks to dies, after which 2D floorplanning is performed
independently on each die and the results are stacked. This improves scalability
but imposes strong structural constraints: early block-to-die decisions are
typically fixed and not revisited during floorplanning. As a result, die assignment
is effectively decoupled from floorplan optimization, which limits joint reasoning
about inter-die connectivity, latency minimization, and area balance. Overall, both categories fall short of fully exploiting the 3D design space,
especially the vertical stacking that is central to 3D ICs.
More importantly, once the die assignment is fixed upfront, performance-critical
inter-die links are difficult to reshape systematically during floorplanning, which
limits latency-aware optimization under F2F bonding and can lead to suboptimal
area balance and connectivity patterns.

To address these limitations, we propose \textsc{Great3D}, a scalable,
latency-aware, partitioning-free framework that directly optimizes a native 3D
floorplan for F2F-bonded ICs, building on our earlier native-3D
work~\cite{2026ASPDAC_Great3D}.
The framework explicitly models inter-die connectivity and jointly
optimizes block placement and die assignment within a single optimization flow.
\revise{Throughout, ``partitioning-free'' refers to the absence of a predefined die-level netlist partitioning step: the framework operates directly on a given block-level netlist and determines die assignment within the optimization itself, with the block-level granularity fixed by the upstream RTL hierarchy.}

Specifically, we first obtain a globally informed 3D embedding using an
SDP-based formulation.
We then introduce a dynamic-programming-based refinement to strengthen
die-assignment decisions under area balance and feasibility constraints.
Finally, the refined solution is further improved through continuous
2D optimization under spacing and fixed-outline constraints.
As illustrated in~\Cref{fig:intra2inter}, by explicitly leveraging inter-die
connections in a 3D stack, our optimization better exploits the vertical
dimension and produces a more compact 3D floorplan under identical outline
constraints.
Such latency-aware, partitioning-free co-optimization not only improves layout
density but also reduces system-level latency, yielding lower cycles per instruction (CPI).

Overall, this work makes the following contributions:
\begin{itemize}
      \item We introduce a native-3D floorplanning framework that jointly optimizes
            block placement and die assignment without relying on a predefined netlist
            partitioning, enabling comprehensive exploration of the 3D design space.

      \item We propose a latency-aware unified objective for F2F-bonded 3D ICs by coupling
            interconnect cost with a CPI-derived latency term to reflect system-level
            performance impact, addressing the non-uniform communication costs in 3D stacks.

      \item We develop a three-stage optimization procedure combining (1)~SDP-based 3D global placement, (2)~dynamic-programming-based die-assignment refinement with explicit area-balance optimization via contiguous-interval search, and (3)~L-BFGS-B continuous refinement under spacing and fixed-outline constraints.

      \item We incorporate comprehensive feasibility handling, including fixed-outline, aspect-ratio, and hybrid-bonding density constraints, ensuring the generated 3D floorplans satisfy practical design rules.

      \item Extensive experiments on the GSRC and ATPlace benchmark suites demonstrate that \textsc{Great3D} delivers strong wirelength quality and favorable overall trade-offs compared with state-of-the-art baselines. \textcolor{blue}{On GSRC, it reduces total wirelength by $2.40$--$2.74\times$ on average over competing 3D floorplanners, and its DP die-assignment stage improves CPI by $9.5$--$17.8\%$.} We provide complexity analysis and ablation studies isolating the contribution of each optimization stage.
\end{itemize}

\noindent The rest of this paper is organized as follows.
\Cref{sec:prelim} introduces the preliminaries and our formulation of the 3D floorplanning problem.
\Cref{sec:method} presents the proposed \textsc{Great3D} framework,
including an overview of the optimization flow (\Cref{subsec:overview}) and the key components:
SDP-based 3D optimization (\Cref{subsec:SDP_optimization}), dynamic programming-based die-assignment refinement (\Cref{subsec:3d22d_dp_optimization}), and the final
2D continuous refinement (\Cref{subsec:2d_refinement}).
\Cref{sec:exp} reports the experimental setup and results.

%% file: doc/2-prelim.tex
\section{Preliminaries}
\label{sec:prelim}

\subsection{3D IC}
\label{subsec:hybrid_bonding_3dic}

Modern 3D integration can be broadly classified into three categories~\cite{2016types-of-3d-ic, 2025ISPD-3DIC-Yu, kim2021microbumping}:
(1) TSV-based integration,
(2) monolithic 3D ICs, and
(3) face-to-face (F2F) hybrid bonding.
TSV-based integration~\cite{2017TCAD-Lu} introduces significant parasitics and area overhead, and is therefore better suited to sparse inter-die connectivity.
Monolithic 3D ICs~\cite{2021DAC-Sai} support extremely fine-grained integration via nanoscale monolithic inter-tier vias (MIVs), but require sequential device processing, which increases fabrication cost and reduces yield.
In contrast, F2F hybrid bonding~\cite{2023ECTC-hybrid-bonding, 2023ECTC-Netzband, 2023ECTC-hybrid-bonding-process} aligns pre-fabricated dies at their top metal layers, providing dense, low-parasitic vertical interconnects.
It naturally supports heterogeneous dies and die reuse, and is increasingly adopted in commercial 3D ICs~\cite{ISPD24-Liu, zhou2024research}.

\textcolor{blue}{The main symbols used throughout are: $N_p$ (per-net connectivity matrix of net $e_p$), $\mathbf{S}=\sum_p N_p$ (structure matrix aggregating all per-net connectivity), $A^o$ (the netlist-derived pairwise \emph{connection matrix}), $A$ (the latency-weighted connection matrix combining $A^o$ and $\mathbf{S}$), $L$ (pairwise latency matrix), and $D$ (the squared-Euclidean-distance surrogate matrix used in the SDP relaxation).}

The proposed \textsc{Great3D} framework primarily targets F2F-bonded 3D systems, and can be extended to multiple dies.
To connect physical layout with system-level performance, we adopt a CPI-based latency model following~\cite{2023ICCAD-Zhuang-3d-bonding}.
\revise{In this notation, ``sta'' denotes a static baseline term, ``lat'' denotes a latency-dependent term, and ``ppi'' follows the original architecture-level notation for the $i$-th package product. In our floorplanning formulation, the ppi-related term is used as the placement-dependent inter-block communication component. Thus, different floorplans of the same design share the same static term, while the ppi-related latency term changes with floorplan-induced wirelength and block-to-block communication distances.}
Let $\mathbf{S}$ denote the structure matrix aggregated from per-net connectivity matrices $N_p$, and $L \in \mathbb{R}^{n \times n}$ the pairwise latency matrix.
The generalized CPI is modeled as
\begin{equation}
    \mathcal{H}(\mathbf{S}, L, \mathrm{MP}) = \text{CPI}_{\text{sta}} + \alpha_1 \cdot (\text{lat}_{\text{ppi}} - \text{lat}_{\text{sta}}),
    \label{eq:3dflp_hybrid-bonding-latency}
\end{equation}
where $\text{lat}_{\text{ppi}} = \langle \mathbf{S}, L \rangle$ is the Frobenius inner product between the structure and latency matrices, capturing path-length–weighted communication latency at the block level.
Here $N_p$ is the connectivity matrix for net $e_p$, and $\mathbf{S} = \sum_p N_p$.
\textcolor{blue}{For example, suppose a net $e_p$ connects blocks $b_1$, $b_2$, and $b_4$. Then $N_p$ has nonzero entries for the pairwise block interactions induced by this net, i.e., $(N_p)_{12}=(N_p)_{21}=(N_p)_{14}=(N_p)_{41}=(N_p)_{24}=(N_p)_{42}=1$, while unrelated entries are zero. Summing all $N_p$ over nets gives $\mathbf{S}$. Therefore, $S_{ij}$ represents the accumulated netlist-induced communication strength between blocks $b_i$ and $b_j$, and serves as the weight of the pairwise latency $L_{ij}$ in $\langle \mathbf{S},L\rangle$.}
The static CPI is modeled as $\text{CPI}_{\text{sta}} = \alpha_2 \cdot \mathrm{MP} + \beta_1$, \textcolor{blue}{while the placement-dependent term $\alpha_1(\text{lat}_{\text{ppi}} - \text{lat}_{\text{sta}})$ in \Cref{eq:3dflp_hybrid-bonding-latency} represents the interconnect delay contribution and is taken proportional to the netlist-level wirelength, i.e., $\alpha_1(\text{lat}_{\text{ppi}} - \text{lat}_{\text{sta}}) = \beta_2 \cdot \text{wirelength}_{\text{ppi}}$.} Here $\mathrm{MP}$ captures system-level microarchitectural parameters (e.g., miss penalties) and $\text{wirelength}_{\text{ppi}}$ is the netlist-level wirelength under the given floorplan.
This model maps 3D floorplanning decisions to a CPI-based performance metric.
\revise{Since the static term $\text{CPI}_{\text{sta}}$ is fixed for a given design, a floorplan that reduces $\text{lat}_{\text{ppi}}$ by shortening latency-critical block-to-block communication paths lowers $\alpha_1(\text{lat}_{\text{ppi}}-\text{lat}_{\text{sta}})$ and hence the generalized CPI; smaller values therefore indicate better performance.}

\subsection{3D Floorplanning}
\label{subsec:3dfloorplan}
In floorplanning, communication cost is commonly estimated using the half-perimeter wirelength (HPWL) model. In 2D, the HPWL of a net $e_i$ connecting a set of blocks $S_i$ is defined in~\Cref{eq:2D_HPWL}, which is widely adopted due to its simplicity and effectiveness~\cite{2023DAC-Li,2008ASPDAC-Luo-AR}.
In practical optimization frameworks, multi-pin nets are commonly decomposed into pairwise interactions to yield differentiable surrogate objectives~\cite{2023DAC-Li}. We adopt such a pairwise formulation, defining a \textcolor{blue}{connection matrix} $A^o$, where each entry $A^{o}_{ij}$ represents the aggregated connection strength between blocks $i$ and $j$. This yields the objective function expressed as the inner product between \textcolor{blue}{connection} and distance matrices as \Cref{eq:wirelength_obj}.
\textcolor{blue}{HPWL is the metric used for wirelength reporting and evaluation, whereas the SDP-based global embedding uses a pairwise squared Euclidean distance surrogate as a continuous relaxation, where $D_{ij}$ denotes the squared Euclidean distance surrogate between blocks $i$ and $j$ used as the continuous relaxation in the SDP global embedding.} 
\begin{equation}
    \mathrm{WL}_{2D}(e_i) = \max_{b \in S_i}(x_b) - \min_{b \in S_i}(x_b)
    + \max_{b \in S_i}(y_b) - \min_{b \in S_i}(y_b),
    \label{eq:2D_HPWL}
\end{equation}
\begin{equation}
    \mathcal{W}(A^o, D) = \langle A^o, D \rangle = \sum_{i,j} A^o_{ij} \cdot D_{ij}.
    \label{eq:wirelength_obj}
\end{equation}
In 3D floorplanning, we naturally extend this geometric interpretation into 3D Euclidean space.
This connection-matrix representation is used in our unified objective in conjunction with \textcolor{blue}{the squared Euclidean distance surrogate}.
\subsection{Problem Formulation}
\label{subsec:formulation}
We now formalize the 3D floorplanning problem addressed in this work by specifying the inputs, outputs, constraints, and high-level objective.

\mysection{Input:}
\begin{itemize}
    \item \emph{Netlist} $(V_b, V_p, E)$:
          $V_b = \{b_1, \dots, b_n\}$ is the set of blocks,
          $V_p = \{p_1, \dots, p_m\}$ is the set of fixed pins,
          and $E = \{e_1, \dots, e_k\}$ is the set of nets, each connecting a subset of blocks and/or pins.
    \item \emph{Block parameters}:
          For each block $b_i$, we are given its width $w_i$, height $h_i$, and area $a_i$.
          Soft blocks must satisfy aspect-ratio bounds
          $r_i^{\min} \leq w_i / h_i \leq r_i^{\max}$.
    \item \emph{Design constraints}:
          The layout is confined to a fixed outline $(W, H)$ per die.
          We allow a small overlap tolerance $\epsilon_{\text{overlap}}$, and optionally specify feasible F2F bonding regions $\mathcal{B}_i$ for certain block pairs or regions.
\end{itemize}

\mysection{Output:}
\begin{itemize}
    \item A 3D floorplan specifying the center coordinate $(x_i, y_i, z_i)$ for each block $b_i$,
          where $(x_i, y_i)$ are continuous in-die coordinates and $z_i$ encodes the discrete die assignment.
          \textcolor{blue}{Equivalently, we collect all block coordinates into a matrix $\mathbf{X}\in\mathbb{R}^{n\times 3}$, where the $i$-th row $\mathbf{X}_i=[x_i,y_i,z_i]$ represents the 3D coordinate of block $b_i$.}
    \item Per-die floorplan dimensions $(W, H)$ and legal block shapes $(w_i, h_i)$ satisfying area and aspect-ratio constraints.
    \item Net partitioning induced by the die assignment:
          $E_{\text{top}}$, $E_{\text{bot}}$, and $E_{\text{inter-die}}$.
    \item Inter-die connection statistics, including the total number of inter-die nets $|E_{\text{inter-die}}|$ and the corresponding density $\rho_{\text{cur}}$.
          \textcolor{blue}{We compute the current density as $\rho_{\text{cur}}=|E_{\text{inter-die}}|/\mathrm{area}_{\text{die}}$, where $\mathrm{area}_{\text{die}}=W\cdot H$ denotes the F2F bonding-interface area.}
\end{itemize}

\mysection{Constraints:}
To ensure manufacturability and geometric feasibility, the 3D floorplan must satisfy the following constraints.
\begin{itemize}
    \item \textbf{Overlap constraint:}
          Blocks must not overlap beyond a prescribed tolerance $\epsilon_{\text{overlap}}$.
          Let $\mathrm{Overlap}(b_i, b_j)$ denote the overlap area between blocks $b_i$ and $b_j$.
          Then
          \begin{equation}
              \sum_{i<j} \mathrm{Overlap}(b_i, b_j) \leq \epsilon_{\text{overlap}}.
              \label{eq:constraint_overlap}
          \end{equation}
    \item \textbf{Aspect-ratio constraint:}
          For soft blocks, the aspect ratio must lie within given bounds:
          \begin{equation}
              r_i^{\min} \leq \frac{w_i}{h_i} \leq r_i^{\max},
              \quad \forall b_i \in B_{\text{soft}}.
              \label{eq:constraint_aspect_ratio}
          \end{equation}
    \item \textbf{Fixed-outline constraint:}
          All blocks must lie within the die outline $(W, H)$.
          In addition, we assume that the top and bottom dies share the same outline:
          \begin{equation}
              \begin{aligned}
                   & 0 \leq x_i - \frac{w_i}{2}, \quad x_i + \frac{w_i}{2} \leq W, \\
                   & 0 \leq y_i - \frac{h_i}{2}, \quad y_i + \frac{h_i}{2} \leq H,
                  \quad \forall b_i \in B.
              \end{aligned}
              \label{eq:constraint_outline}
          \end{equation}
    \item \textbf{Hybrid-bonding density constraint:}
          To reflect practical F2F bonding limits, we bound the density of vertical connections:
          \begin{equation}
              |E_{\text{inter-die}}| \leq \rho_{\text{max}} \cdot \text{area}_{\text{die}},
              \label{eq:constraint_bonding_density}
          \end{equation}
          where $\rho_{\text{max}}$ is the maximum allowable inter-die connection density dictated by the bonding technology.
          \textcolor{blue}{Equivalently, the constraint requires $\rho_{\text{cur}}\leq\rho_{\text{max}}$.}
          \revise{This constraint is a global floorplanning-level abstraction; local bonding-interface density is resolved by finer-grained bump/pad assignment and routing-level verification.}
\end{itemize}

\begin{figure*}[t]
    \centering
    \hspace*{-0.3cm}  %
    \includegraphics[width=\linewidth]{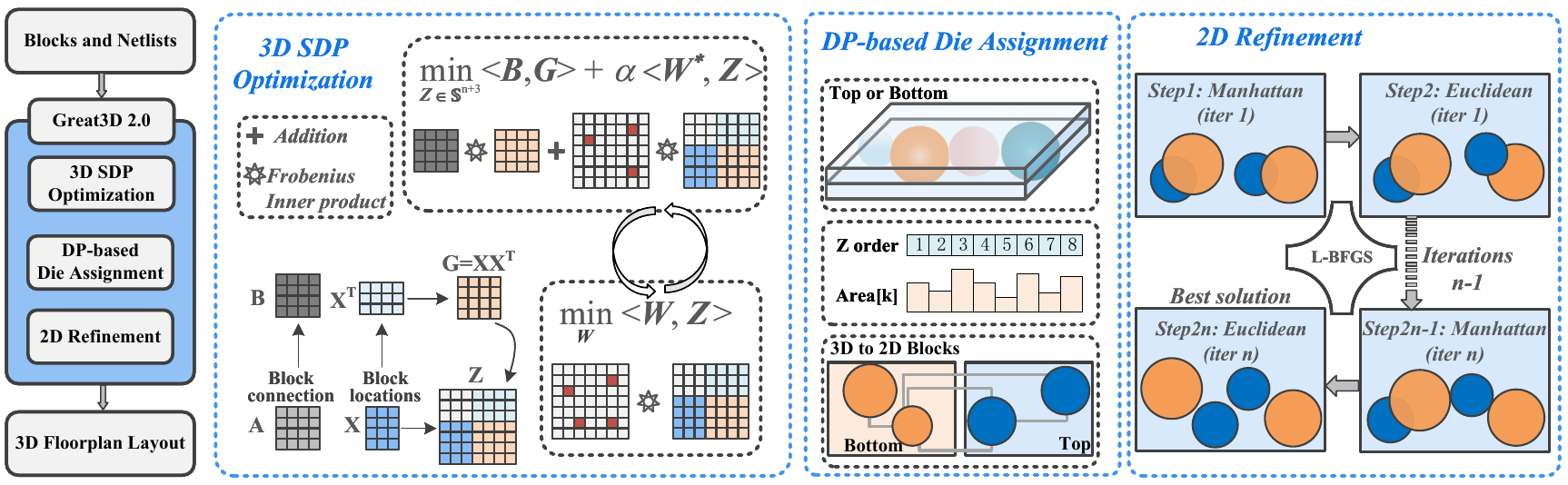}
    \caption{Overview of the proposed \textsc{Great3D} framework in~\Cref{subsec:overview}.
        \textsc{Great3D} follows a three-stage optimization flow: a global 3D SDP embedding
        (\Cref{subsec:SDP_optimization}), a DP-based die assignment refinement that exploits the SDP-induced
        $z$-ordering (\Cref{subsec:3d22d_dp_optimization}), and an intra-die 2D refinement using
        L-BFGS-B (\Cref{subsec:2d_refinement}).}

    \label{fig:framework}
\end{figure*}
\mysection{Objective:}
Given the above inputs and constraints, our goal is to co-optimize traditional floorplan quality metrics and 3D-integration-specific performance.
At a high level, we consider both the total wirelength and the CPI-based latency model, and formulate a direct trade-off:
\begin{equation}
    \min_{\Theta} \quad
    \mathcal{F} = (1 - \mu) \cdot \mathcal{W}(E)
    + \mu \cdot \mathcal{H}(\mathbf{S}, L, \mathrm{MP}),
    \label{eq:objective_function}
\end{equation}
where $\mathcal{W}(\cdot)$ denotes total wirelength (e.g., summed 3D HPWL) and $\mathcal{H}(\cdot)$ is the latency cost in~\eqref{eq:3dflp_hybrid-bonding-latency}.
The variable set $\Theta$ includes all block coordinates and die assignments, and \textcolor{blue}{the scalar $\mu \in [0,1]$ balances the wirelength-driven placement objective and the CPI-based communication-latency objective. Here, $\mathcal{W}(E)$ is the HPWL-based wirelength objective introduced in \Cref{eq:2D_HPWL,eq:wirelength_obj}; in floorplanning, reducing this term encourages connected blocks to be placed closer to each other and therefore serves as a standard proxy for compact wirelength-driven placement quality.}

Directly optimizing the heterogeneous terms $\mathcal{W}(E)$ and $\mathcal{H}(\mathbf{S}, L, \mathrm{MP})$ is challenging, since they are defined over different domains.
To enable a unified optimization, we decompose the second term and introduce weighted components.
We first rewrite the objective as
\begin{equation}
    \mathcal{F} =
    (1 - \mu) \cdot \sum_{i,j} A^{o}_{ij} \cdot d_{ij}
    + \mu \cdot \sum_{i,j} S_{ij} \cdot L_{ij},
    \label{eq:f1_f2_spread}
\end{equation}
where $A^{o}$ is the original connection matrix derived from the netlist, \textcolor{blue}{$d_{ij}$ is the squared Euclidean distance surrogate between blocks $i$ and $j$}, and $S_{ij}$ and $L_{ij}$ are the entries of the structure and latency matrices, respectively.
We then define a weighted connection matrix
\begin{equation}
    A_{ij} = (1-\mu)\cdot A^{o}_{ij} + \mu \cdot S_{ij} \cdot \lambda_{ij},
\end{equation}
where $\lambda_{ij}$ absorbs the latency-related scaling\textcolor{blue}{, so that the pairwise block-level latency is realized through the placement distance as $L_{ij} = \lambda_{ij}\,d_{ij}$; substituting this relation, the latency term $\sum_{i,j} S_{ij} L_{ij}$ in \Cref{eq:f1_f2_spread} becomes $\sum_{i,j} S_{ij} \lambda_{ij} d_{ij}$, which matches the second term of the unified objective in \Cref{eq:unified_obj}}.
\textcolor{blue}{The first term preserves the original netlist connection strength, while the second term introduces latency-aware connection strength. Here, $\lambda_{ij}$ denotes the pairwise latency cost between blocks $i$ and $j$. Thus, $S_{ij}\lambda_{ij}$ assigns a larger weight to block pairs that are both strongly connected and latency-critical. Given a design, we compute $A^o$ from the netlist, construct $S$ from the per-net connectivity matrices, obtain $\lambda_{ij}$ from the pairwise latency model, and then evaluate $A_{ij}$ using the above equation. In this matrix form, $\mu$ controls the relative contribution of the original wirelength-driven edge weight and the latency-aware edge weight in the unified matrix $A$.}
\revise{This conversion links the CPI notation to the floorplanning objective: the static CPI component remains fixed for a given design, while the placement-dependent latency component is represented by pairwise communication weights in $A$. The subsequent SDP-based optimization then minimizes the weighted block-to-block distances under the 3D floorplanning constraints.}
Using this unified matrix and the distance matrix $D$ with entries $D_{ij} = d_{ij}$, the objective can be compactly expressed as
\begin{equation}
    \mathcal{F} = \sum_{i,j} A_{ij} \cdot d_{ij} = \langle A, D \rangle.
    \label{eq:unified_obj}
\end{equation}

This formulation converts the original multi-objective problem into a single weighted connection–distance matching problem, and provides the foundation for the SDP-based 3D embedding and subsequent refinement described in Section~\ref{sec:method}.

%% file: doc/3-Methodology.tex
\section{Methodology of \shorttitle}
\label{sec:method}

\subsection{Framework Overview}
\label{subsec:overview}

As shown in~\Cref{fig:framework}, \shorttitle{} follows a three-stage
optimization flow that operates directly in a native 3D solution space.
\revise{For example, consider four soft blocks $b_1$--$b_4$, where $b_1$ and $b_2$ are strongly connected and $b_3$ and $b_4$ form another strongly connected pair.
The SDP embedding first places these blocks in a continuous 3D space, so that blocks with stronger connection weights tend to be closer and their $z$-coordinates provide a global indication of die affinity.
Sorting the blocks by the SDP-induced $z$-coordinates then gives an ordered sequence for die assignment.
Instead of simply cutting this sequence by block count, the DP refinement searches feasible contiguous intervals whose total area better matches the die-area target while avoiding unnecessary splits of strongly connected groups.
Finally, after the die assignment is fixed, the L-BFGS-B-based 2D refinement optimizes the in-plane coordinates on each die to reduce wirelength and satisfy spacing constraints.}
Unlike partition-first approaches that commit to die assignments before
floorplanning, \shorttitle{} jointly optimizes block floorplanning and die
assignment within a unified analytical framework, enabling joint reasoning
about inter-die connectivity and intra-die placement quality.
Beyond a baseline two-stage design with a simple median cut for die
assignment~\cite{2026ASPDAC_Great3D}, \shorttitle{} employs a dynamic-programming
refinement stage that explicitly balances die area while preserving spatial
coherence (\Cref{subsec:3d22d_dp_optimization}).

Concretely, the SDP stage (\Cref{subsec:SDP_optimization}) produces a soft 3D
embedding $\mathbf{X} \in \mathbb{R}^{n \times 3}$ whose $z$-ordering gives a
globally informed indication of die affinity, with strongly connected blocks
sharing similar $z$-coordinates; the DP stage
(\Cref{subsec:3d22d_dp_optimization}) turns this ordering into an area-balanced
die assignment; and the L-BFGS-B stage (\Cref{subsec:2d_refinement}) refines the
in-plane positions on each die under spacing and outline constraints, bridging
the soft-block SDP embedding to the final per-die floorplan.

\subsection{3D SDP Optimization}
\label{subsec:SDP_optimization}
This stage produces an initial 3D embedding of all blocks by solving a semidefinite programming (SDP) relaxation. The key idea is to treat wirelength minimization as a graph embedding problem: connected blocks should be placed close together in 3D space. The SDP formulation provides a \textcolor{blue}{tractable convex relaxation} of this embedding, and the resulting $z$-coordinates naturally indicate which blocks should share the same die.

Following the notations in Section~\ref{sec:prelim}, the Gram matrix is defined as $\mathbf{G} = \mathbf{X} \mathbf{X}^{\top}$. Thus, the objective function is rewritten as ~\Cref{equa::ad2ag}.
\begin{equation}
    \langle \mathbf{A}, \mathbf{D} \rangle = \sum_{i=1}^{n} \sum_{j=1}^{n} \mathbf{A}_{ij} \left( \mathbf{G}_{ii} + \mathbf{G}_{jj} - 2\mathbf{G}_{ij} \right).
    \label{equa::ad2ag}
\end{equation}
Following~\cite{2023DAC-Li}, we construct a matrix $\mathbf{B}$ to simplify the calculation of the matrix according to the mathematical transformation in~\Cref{equa::b_def}. This definition clearly distinguishes the diagonal elements (which involve aggregated in- and out-connections of vertex \(i\)) from the off-diagonal ones (which are simply scaled by a constant). The $\bf{Z}$ is defined as \Cref{eq:Z_def}
\begin{equation}
    \mathbf{B}_{ij} =
    \begin{cases}
        \displaystyle \sum_{k=1}^{n} \mathbf{A}_{ik} + \sum_{k=1}^{n} \mathbf{A}_{kj}, & \text{if } i = j,    \\[5pt]
        -2\, \mathbf{A}_{ij},                                                          & \text{if } i \neq j.
    \end{cases}
    \label{equa::b_def}
\end{equation}

\begin{equation}
    \bm{Z} =
    \begin{bmatrix}
        \bm{I}_{3 \times 3} & \bm{X}^{\top}_{3 \times n} \\
        \bm{X}_{n \times 3} & \bm{G}_{n \times n}
    \end{bmatrix}
    \in \mathbb{S}_{+}^{3 + n}.
    \label{eq:Z_def}
\end{equation}
where \( \bm{I}_{3 \times 3} \) is the identity block and \( \bm{G}_{n \times n}\succeq 0 \) is the symmetric Gram block; the construction requires a rank constraint, explained in~\Cref{eq:constraint2_bounds}. The top-3 eigenvectors of $\mathbf{Z}$, by the Courant-Fischer principle, capture the dominant spatial directions for recovering the 3D coordinates.
The optimization alternates between two coupled subproblems, \Cref{eq:update_Z,eq:update_W}, at each iteration $k$.
\begin{equation}
    \mathbf{Z}^{(k+1)} = \mathcal{P}_{1}\bigl(\mathbf{Z}^{(k)}, \mathbf{W}^{(k)}; f_{1}, \mathcal{C}_{1}\bigr),
    \label{eq:update_Z}
\end{equation}
\begin{equation}
    \mathbf{W}^{(k+1)} = \mathcal{P}_{2}\bigl(\mathbf{W}^{(k)}, \mathbf{Z}^{(k+1)}; f_{2}, \mathcal{C}_{2}\bigr),
    \label{eq:update_W}
\end{equation}
\begin{equation}
    f_1(\mathbf{Z}) = \langle \mathbf{B}, \mathbf{G}(\mathbf{Z}) \rangle + \alpha \langle \mathbf{W}^*, \mathbf{Z} \rangle,
    \label{eq:objective_f1}
\end{equation}
\begin{equation}
    f_2(\mathbf{W}) = \langle \mathbf{W}, \mathbf{Z}^* \rangle.
    \label{eq:objective_f2}
\end{equation}
where $f_1,f_2$ are the objectives and $\mathcal{C}_1,\mathcal{C}_2$ the associated constraint sets. In Subproblem~1 (\Cref{eq:objective_f1}), $\mathbf{G}(\mathbf{Z})$ encodes the placement-induced distances. \textcolor{blue}{Following the rank-constraint relaxation in SDP-based global floorplanning~\cite{2023DAC-Li}, the second term uses a fixed direction matrix $\mathbf{W}^*$ (controlled by $\alpha$) to regularize toward the rank-3 embedding desired for $\mathbf{Z}\in\mathbb{S}_{+}^{n+3}$, penalizing components outside the dominant 3D subspace; $\mathbf{W}^*$ is the current value obtained from the $\mathbf{W}$-subproblem rather than a manually specified matrix.} Subproblem~2 refines $\mathbf{W}$ by aligning it with the current solution $\mathbf{Z}^*$. Non-overlap requires $\mathbf{Z}$ to satisfy~\Cref{eq:constraint1_nonoverlap}.
\begin{equation}
    D_{ij} = \mathbf{Z}_{ii} + \mathbf{Z}_{jj} - 2\mathbf{Z}_{ij} \geq (r_i + r_j)^2,\quad \forall i \neq j,\quad \mathbf{Z} \succeq 0,
    \label{eq:constraint1_nonoverlap}
\end{equation}
\begin{equation}
    \mathbf{0} \preceq \mathbf{W} \preceq \mathbf{I},\quad \mathrm{Trace}(\mathbf{W}) = n
    \label{eq:constraint2_bounds}
\end{equation}
which ensures that any two blocks are separated by at least the sum of their radii and that $\mathbf{Z}$ remains positive semidefinite. The dual variable $\mathbf{W}$ is constrained to the normalized semidefinite cone of \Cref{eq:constraint2_bounds}, keeping its eigenvalues in $(0,1)$ with normalized spectral weight.
\textcolor{blue}{Equivalently, the $\mathbf{W}$-update selects the $n$-dimensional subspace associated with the smallest eigenvalues of the current $\mathbf{Z}^*$, leaving the top three directions for the recovered 3D coordinates. If $\mathbf{u}_1,\ldots,\mathbf{u}_n$ denote the corresponding normalized eigenvectors, the updated direction matrix can be expressed as $\mathbf{W}^*=\sum_{\ell=1}^{n}\mathbf{u}_{\ell}\mathbf{u}_{\ell}^{\top}$. The regularization term $\langle\mathbf{W}^*,\mathbf{Z}\rangle$ therefore suppresses energy in the undesired higher-rank subspace and guides the SDP solution toward a rank-3 embedding.}

After solving the SDP, we recover the 3D layout by spectral decomposition of the
Gram matrix $\mathbf{G}=\mathbf{V}\mathbf{\Lambda}\mathbf{V}^{\top}$.
The top-3 eigenvectors $\mathbf{V}_3\in\mathbb{R}^{n\times 3}$, scaled by the
square roots of their eigenvalues, yield the recovered coordinates
$\mathbf{X}=\mathbf{V}_3\mathbf{\Sigma}_{\text{sqrt}}$. The resulting
$z$-coordinates (the third column of $\mathbf{X}$) induce an ordered sequence of
modules along the vertical axis, which will be exploited by the DP-based die
assignment refinement in~\Cref{subsec:3d22d_dp_optimization}.
Because the downstream DP stage mainly leverages the induced $z$-ordering as a coarse die-affinity signal, an approximate SDP solution is sufficient in practice; we terminate the SDP solver once the recovered $z$-ordering stabilizes and delegate fine-grained optimization to the subsequent refinement stages.

\mysection{Complexity analysis:}
The SDP relaxation can be solved using interior-point methods with $O(n^3)$ time
per iteration, where typical convergence requires $O(\sqrt{n})$ iterations,
yielding overall $O(n^{3.5})$ complexity. The spectral decomposition for coordinate
recovery incurs $O(n^3)$ cost. \textcolor{blue}{For the block-level 3D floorplanning instances targeted here ($n\le 500$), this cubic-order cost remains tractable.}

\subsection{DP-based Die Assignment Refinement}
\label{subsec:3d22d_dp_optimization}
This stage converts the continuous SDP solution into a discrete die assignment by selecting an area-balanced contiguous interval along the SDP-induced $z$-ordering. The goal is to partition blocks into top and bottom dies such that the total area on each die is balanced, while keeping strongly connected blocks together.

The SDP embedding described in~\Cref{subsec:SDP_optimization} produces
soft spheres with an inherent ordering along the $z$-axis of the
three-dimensional layout space. This ordering reflects the global
connectivity structure: blocks that are strongly connected tend to have
similar $z$-coordinates, indicating they should be placed on the same die
to minimize inter-die communication. However, the SDP solution is only a
continuous relaxation and does not directly enforce discrete die assignment
or area balance constraints.

A naïve yet effective baseline~\cite{2026ASPDAC_Great3D}
simply cuts this sequence at its midpoint: blocks whose $z$-coordinates lie
below the median are assigned to the bottom die, and those above the median
to the top die. Although this median-cut strategy already performs reasonably well
in our evaluation (see \Cref{subsec:cmp_results}), it has two significant
drawbacks that limit solution quality:
\begin{itemize}[leftmargin=*, itemsep=2pt, topsep=2pt]
    \item \textbf{No explicit area balance:} The median cut bisects the \emph{number} of blocks, but because blocks have heterogeneous areas, the resulting dies may have vastly different total areas, potentially violating fixed-outline constraints or leaving significant whitespace imbalance.
    \item \textbf{Arbitrary cluster splitting:} Strongly connected clusters that happen to straddle the median $z$-value are split arbitrarily, increasing inter-die communication cost. The median cut does not consider connectivity structure when choosing the split point.
\end{itemize}
\input{algorithms/optimal_subsequence}

A closer examination of the $z$-distribution reveals that we can exploit this ordering more
intelligently: by searching for an optimal contiguous interval rather than a fixed midpoint, we achieve a better trade-off between die area balance and spatial coherence.

To address these limitations, we employ the dynamic-programming procedure in
Algorithm~\ref{alg:dp-split}, which preserves the SDP-induced $z$-ordering for
spatial coherence but chooses the cut to minimize deviation from a target die
area rather than bisecting the sequence. It first sorts blocks by their
$z$-coordinates and builds a prefix-sum table $P$ over the ordered areas $Ar$,
enabling $O(1)$ area queries for any contiguous interval. The target die area is
\begin{equation}
    T = \Bigl(\sum_i Ar[i]\Bigr)\cdot w/2,
    \label{eq:target_area}
\end{equation}
where the whitespace factor $w$ reserves a uniform whitespace fraction per die,
and the divisor $2$ reflects that the two F2F-bonded dies share the same outline
and should therefore have approximately equal total area.

Given the target area $T$, the die assignment problem is formulated as
the following contiguous-interval optimization:
\begin{equation}
    (s^{*}, e^{*})
    =\arg\min_{1 \le s \le e \le n}
    \left| \sum_{k=s}^{e} Ar[k] - T \right|,
    \label{eq:dp_formulation}
\end{equation}
\begin{figure}[t!]
    \centering
    \includegraphics[width=0.99\linewidth]{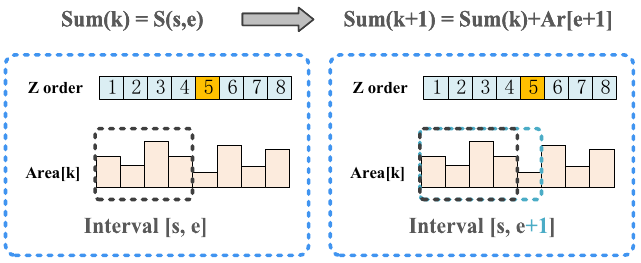}
    \caption{Illustration of the DP-based die assignment refinement.
        Given the SDP-induced $z$-ordering of blocks, ~\Cref{alg:dp-split} searches
        for an optimized interval whose total area best
        matches the target $T$, ensuring balanced die utilization while
        preserving spatial coherence. \textcolor{blue}{This example visualizes the case where the interval starts at $s=1$; in the algorithm, $s$ and $e$ denote the start and end indices of a candidate interval, and all intervals satisfying $1\leq s\leq e\leq n$ are enumerated.}}
    \label{fig:dp_illustration}
\end{figure}
where the indices are ordered by the SDP-induced $z$-coordinates, and the selected interval $(s^*, e^*)$ corresponds to the set of blocks assigned to the bottom die, while all remaining blocks are placed on the top die.
\revise{The candidate intervals in \Cref{eq:dp_formulation} are defined on the SDP-induced $z$-ordering, which already encodes the weighted interconnect and latency-aware objective; the area-balance criterion thus selects among connectivity-coherent intervals, and the subsequent 2D refinement re-optimizes the actual in-plane coordinates on each die. Although the contiguous-interval split does not add an explicit inter-die interconnect term, the connectivity-aware $z$-ordering keeps strongly coupled blocks at adjacent $z$-coordinates, so a single interval rarely severs such groups; the resulting assignment attains both the lowest total wirelength and the best die-area balance in \Cref{tab:interdie_f2f}, confirming that the interconnect cost is effectively contained rather than merely deferred.}
To efficiently evaluate the objective in~\Cref{eq:dp_formulation}, we define the partial sum:
\begin{equation}
    S(s,e)=\sum_{k=s}^{e} Ar[k],
    \label{eq:dp_partial_sum}
\end{equation}
which represents the total area of blocks from index $s$ to $e$. The recurrence relation
\begin{equation}
    S(s,e+1)=S(s,e)+Ar[e+1]
    \label{eq:dp_recurrence}
\end{equation}
shows that $S(s,e)$ can be computed incrementally. By precomputing a prefix-sum table, any interval sum can be retrieved in $O(1)$ time, enabling efficient evaluation of all candidate intervals.
~\Cref{fig:dp_illustration} visualizes this selection over the SDP-induced ordering (the illustrated interval uses $s=1$ only for clarity; the algorithm searches all feasible $(s,e)$). A nested scan over every contiguous interval $(s,e)$, $1\le s\le e\le n$, selects the one whose accumulated area is closest to $T$; these blocks form the bottom die and the rest the top die, preserving the $z$-ordering.
Unlike the median cut, this explicitly balances die area while inheriting the spatial coherence of the SDP stage (blocks with similar $z$-coordinates stay together), avoiding the inter-die shuffling caused by arbitrary block exchanges. The scan enumerates all $\frac{n(n+1)}{2}$ contiguous intervals exactly once, so it returns the globally area-optimal interval $(s^*,e^*)$ minimizing $|S(s,e)-T|$.

\begin{figure}[t!]
    \centering
    \includegraphics[width=1\linewidth]{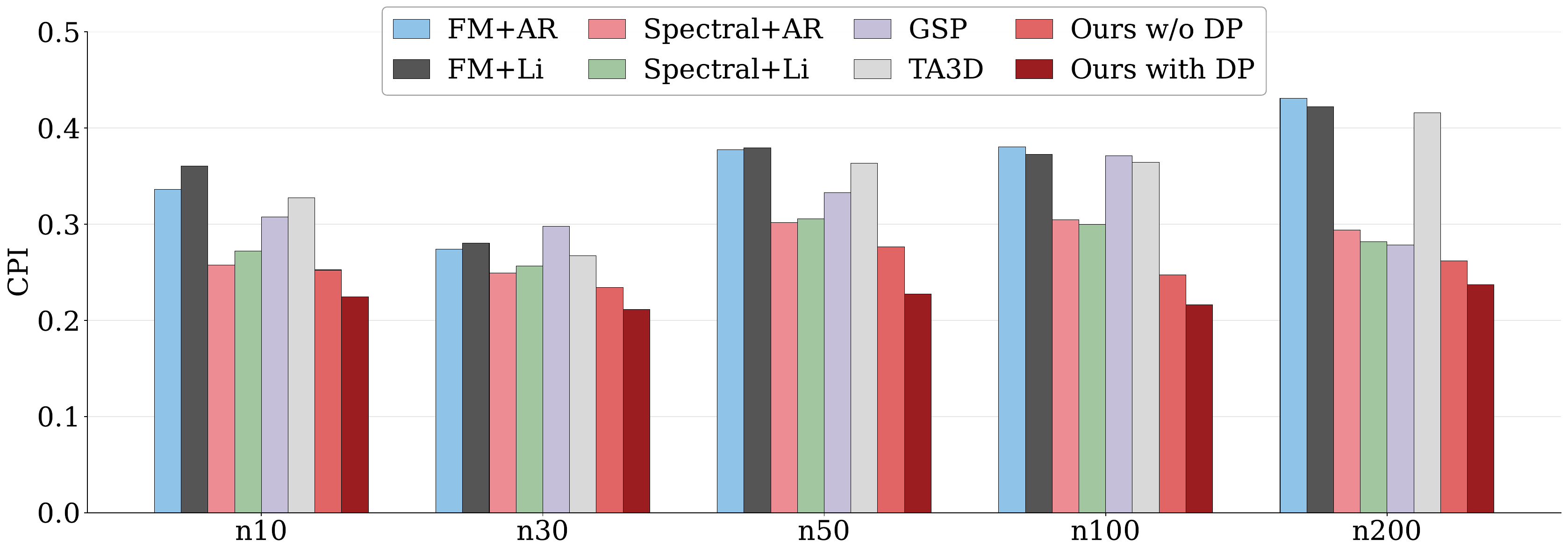}
    \caption{Cycle-per-instruction (CPI) comparison on the GSRC benchmarks \textcolor{blue}{(lower is better; bars are not normalized to our method)}. \textcolor{blue}{``Ours with DP'' is the full \textsc{Great3D}; ``Ours w/o DP'' ablates the DP die-assignment stage.}}
    \label{fig:cpi_comparison}
\end{figure}

\textcolor{blue}{Because the connectivity-aware SDP $z$-ordering already keeps strongly connected blocks at similar $z$-coordinates, a single contiguous-interval split rarely severs such groups; the inter-die-wirelength column of \Cref{tab:interdie_f2f} and the CPI results in \Cref{fig:cpi_comparison},~\Cref{tab:full_comparison} show that the resulting assignment keeps inter-die connectivity balanced and latency low.}

\subsection{2D Refinement}
\label{subsec:2d_refinement}
After die assignment, each die contains a set of blocks with approximate 2D positions inherited from the SDP stage. This stage refines these positions to minimize wirelength while ensuring blocks do not overlap and remain within the die boundary. We formulate this as a continuous optimization problem and solve it using L-BFGS-B, a quasi-Newton method that efficiently handles bound constraints.
\revise{This intra-die refinement is an analytical bridge that turns the continuous 3D SDP centers and the DP die membership into a legal per-die floorplan with explicit spacing and outline constraints, yielding a high-quality initial solution for downstream placement, routing, and signoff. Because projecting the continuous 3D embedding onto a discrete die does not by itself guarantee non-overlap, this stage reduces the residual overlaps.}

Operating within each die separately, Algorithm~\ref{alg:lbfgs-refine} starts from the projected 2D coordinates $\mathbf{X}_0$ of the SDP solution and minimizes the composite objective $f(\mathbf{X})$ of \Cref{eq:lbfgs_obj} (wirelength plus pairwise soft-spacing terms): each iteration computes the gradient, obtains a descent direction via the L-BFGS two-loop recursion~\cite{lbfgs_proposed,lbfgsb_proposed}, takes an Armijo backtracking step~\cite{nocedal2006numerical}, and updates the history pair; the penalty coefficient $\beta$ and norm-mixing ratio $\alpha_3$ are adapted when the update condition is met, until convergence yields the final intra-die placement $\mathbf{X}^*$.

\begin{figure}[t!]
    \centering
    \subfloat[]{ \includegraphics[height=.283\linewidth]{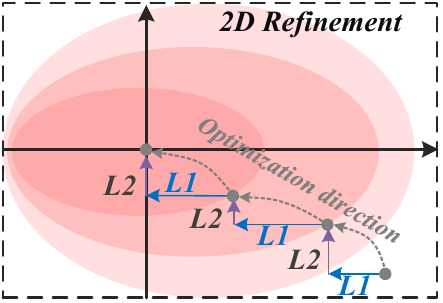} \label{fig:red-a} } \hspace{-.05in}
    \subfloat[]{ \includegraphics[height=.283\linewidth]{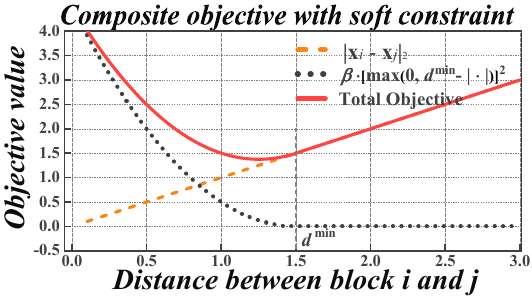} \label{fig:red-b} }
    \caption{
        Illustration of the 2D refinement strategy.
        (a) Alternating \textit{L1} and \textit{L2} gradients improve convergence.
        (b) Soft spacing penalty maintains minimum block separation requirements.
    }
    \label{fig:red}
\end{figure}

After the SDP in ~\Cref{subsec:SDP_optimization}, the objective function in this stage should composite wirelength objective augmented with pairwise soft-spacing constraints, so it is defined as ~\Cref{eq:lbfgs_obj}.
\begin{equation}
    \label{eq:lbfgs_obj}
    f(\mathbf{x}) =
    \sum_{i<j} \mathbf{A}_{ij} \cdot \phi(\mathbf{x}_i, \mathbf{x}_j)
    + \beta\, f_{\mathrm{pen}}(\mathbf{x}).
\end{equation}
\begin{equation}
    \label{eq:comp_phi}
    \phi(\mathbf{x}_i, \mathbf{x}_j) =
    \alpha_3 \|\mathbf{x}_i - \mathbf{x}_j\|_1 +
    (1 - \alpha_3)\|\mathbf{x}_i - \mathbf{x}_j\|_2,
\end{equation}
\begin{equation}
    \label{eq:comp_obj_penalty}
    f_{\mathrm{pen}}(\mathbf{x}) =
    \sum_{i<j}
    \left[
        \max\left(0,\,d_{ij}^{\min} - \|\mathbf{x}_i - \mathbf{x}_j\|_2\right)
        \right]^2.
\end{equation}
where $\mathbf{x}_i = [x_i, y_i]^{\mathsf{T}}$ denotes the 2D position of block $i$ acquired from the procedure in ~\Cref{subsec:SDP_optimization} for each die, and $\beta > 0$ is a regularization weight controlling the strength of spacing penalties. To enhance convergence and escape poor local minima, we adopt a norm-alternating strategy within the potential function ~\Cref{eq:comp_phi}, where a scalar $\alpha_3 \in [0, 1]$ balances Manhattan and Euclidean terms. This composite form leverages the strengths of both norms, as previously shown effective in~\cite{2023DAC-Li}. To maintain legal spacing between block pairs, we incorporate a soft penalty term $f_{\mathrm{pen}}$ defined as ~\Cref{eq:comp_obj_penalty}, where $d_{ij}^{\min} > 0$ denotes the minimum block-center spacing between blocks $i$ and $j$ derived from the geometric constraints in Section~\ref{subsec:formulation}. \textcolor{blue}{Each soft block is modeled as an occupancy region of radius $r_i$ determined from its area, so two blocks are legally separated when their center distance $\|\mathbf{x}_i-\mathbf{x}_j\|_2\ge r_i+r_j$. We take this threshold as $d_{ij}^{\min}$ (unless a larger spacing margin is required), which enforces the surrogate non-overlap constraint $D_{ij}\ge(r_i+r_j)^2$ of \Cref{eq:constraint1_nonoverlap} in the 2D stage.} As illustrated in the right portion of~\Cref{fig:red-b}, this quadratic penalty softly discourages overlaps without introducing hard constraints. The left portion of~\Cref{fig:red-a} visualizes our alternating-norm descent strategy, where L1 and L2 gradients are switched iteratively to improve search direction quality during optimization. With the objective function defined in Eq.~\eqref{eq:lbfgs_obj}, we compute gradients as \Cref{eq:comp_grad}.
\input{algorithms/lbfgs}
\begin{equation}
    \label{eq:comp_grad}
    \frac{\partial f}{\partial \mathbf{x}_i} =
    \sum_{j \ne i} \mathbf{A}_{ij} \cdot \nabla_i \phi(\mathbf{x}_i, \mathbf{x}_j)
    -
    2\beta \sum_{j \ne i}
    \Delta_{ij} \cdot
    \frac{\mathbf{x}_i - \mathbf{x}_j}{\|\mathbf{x}_i - \mathbf{x}_j\|_2},
\end{equation}
\begin{equation}
    \label{eq:phi_grad}
    \nabla_i \phi(\mathbf{x}_i, \mathbf{x}_j) =
    \alpha_3 \cdot \operatorname{sign}(\mathbf{x}_i - \mathbf{x}_j)
    + (1 - \alpha_3) \cdot
    \frac{\mathbf{x}_i - \mathbf{x}_j}{\|\mathbf{x}_i - \mathbf{x}_j\|_2},
\end{equation}
\begin{equation}
    \label{eq:delta_ij}
    \Delta_{ij} = \max\left(0,\,d_{ij}^{\min} - \|\mathbf{x}_i - \mathbf{x}_j\|_2\right).
\end{equation}

Then, we define the update procedure shown in \Cref{eq:two_loop}, where $H_k$ is the L-BFGS approximation of the inverse Hessian, computed using the two-loop recursion described in~\cite{lbfgs_proposed}. This procedure recursively updates the direction using a limited memory of past updates.
\begin{equation}
    \label{eq:two_loop}
    \mathbf{p} = -H_k \nabla f(\mathbf{X}_k),
\end{equation}

Next, the Armijo backtracking line search~\cite{nocedal2006numerical} to compute a valid step size $\eta$ is shown by~\Cref{eq:armijo_condition}. The step is accepted only if it satisfies the sufficient decrease condition (line 6-7):
\begin{equation}
    f(\mathbf{X} + \eta\,\mathbf{p}) \leq f(\mathbf{X}) + c \cdot \eta \cdot \nabla f(\mathbf{X})^\top \mathbf{p}.
    \label{eq:armijo_condition}
\end{equation}
\begin{equation}
    s_k = \mathbf{X}_{k} - \mathbf{X}_{k-1}, \quad
    y_k = \nabla f(\mathbf{X}_{k}) - \nabla f(\mathbf{X}_{k-1}).
    \label{eq:lbfgs_s_y_pair}
\end{equation}
where $c \in (0,1)$ is a small constant (e.g., $10^{-4}$). The optimizer then updates the block coordinates via $\mathbf{X} \gets \mathbf{X} + \eta\,\mathbf{p}$. After each update, the function \textsc{UpdateHistory} appends the new $(s_k, y_k)$ pair. And we discards the oldest pair if the memory limit $m$ is exceeded. These updates (line 8) enable L-BFGS to efficiently approximate second-order curvature without storing full Hessians. Finally, we monitor a set of dynamic refinement (line 9-11) via \textsc{ParameterUpdateCondition}$(\mathbf{X})$, and adapt hyperparameters accordingly. Specifically, if any pairwise spacing violation exceeds a threshold, the penalty weight $\beta$ is increased to enforce tighter spacing control. Similarly, the L1-L2 mix coefficient $\alpha_3$ is adjusted periodically to enhance convergence. The optimization loop continues until convergence criteria are met, such as a small gradient norm or a maximum number of iterations. The refined layout $\mathbf{X}^{*}$ is returned as the output for Great3D framework, which is also the solution of the unified objective shown as ~\Cref{eq:unified_obj} for the 3D IC floorplanning problem proposed in ~\Cref{subsec:formulation}.

\mysection{Complexity analysis:}
The L-BFGS-B method requires $O(mn)$ operations per iteration, where $m \approx 10$ is the memory depth and $n$ is the number of blocks. Each iteration involves two-loop recursion over the stored correction pairs, gradient computation ($O(n^2)$ for pairwise wirelength terms), and Armijo line search. In practice, convergence is typically achieved within 100--200 iterations for instances up to $n=200$, maintaining overall scalability comparable to classical 2D placement algorithms.

\begin{figure*}[t!]
    \centering
    \begin{minipage}{0.32\textwidth}
        \includegraphics[width=\linewidth]{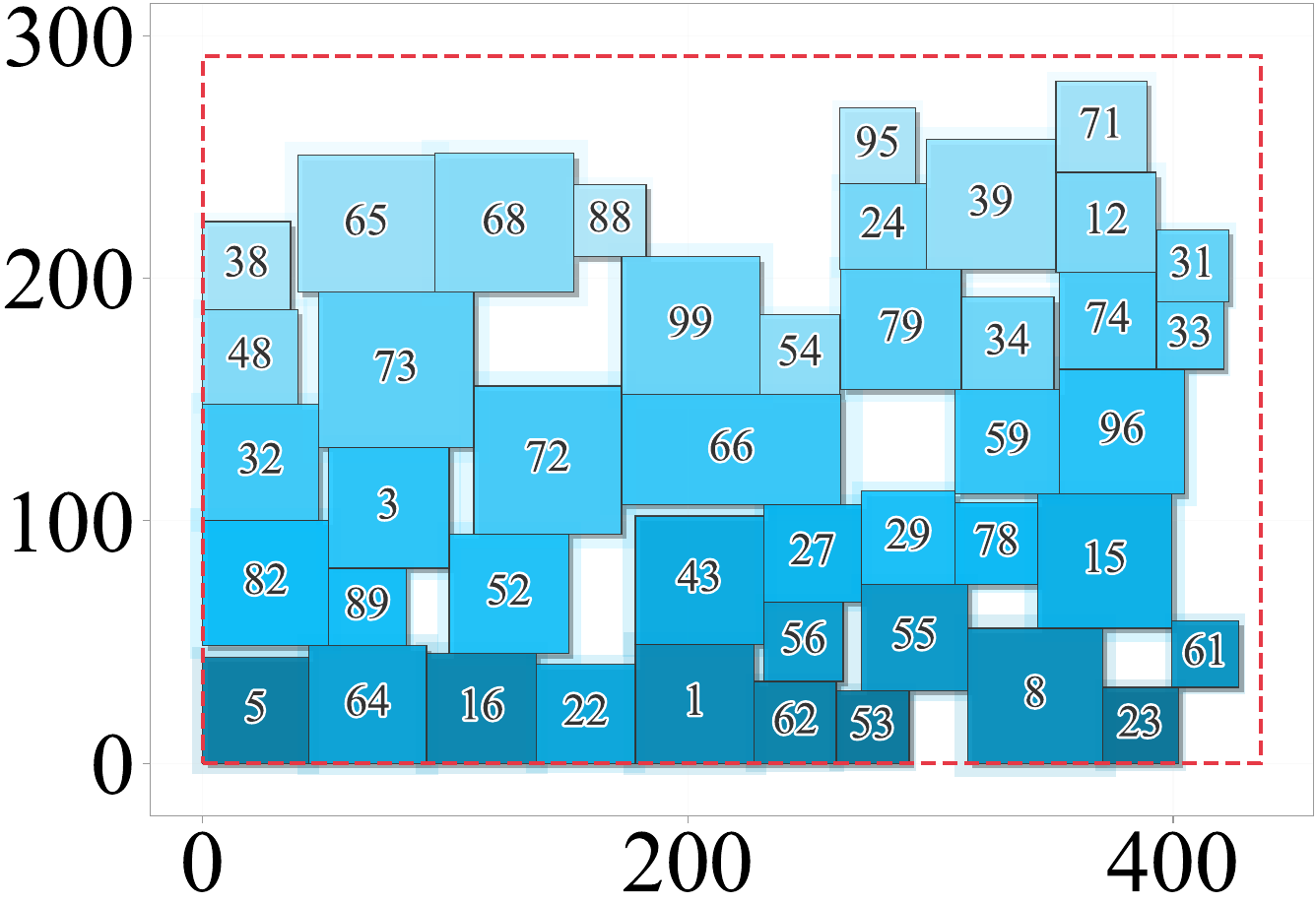}
        \captionof{figure}{Top die projection of the legalized 3D floorplan (aspect ratio 4:3); the red dashed outline is the fixed die boundary.}
        \label{fig:top_die_floorplan}
    \end{minipage}
    \hfill
    \begin{minipage}{0.32\textwidth}
        \includegraphics[width=\linewidth]{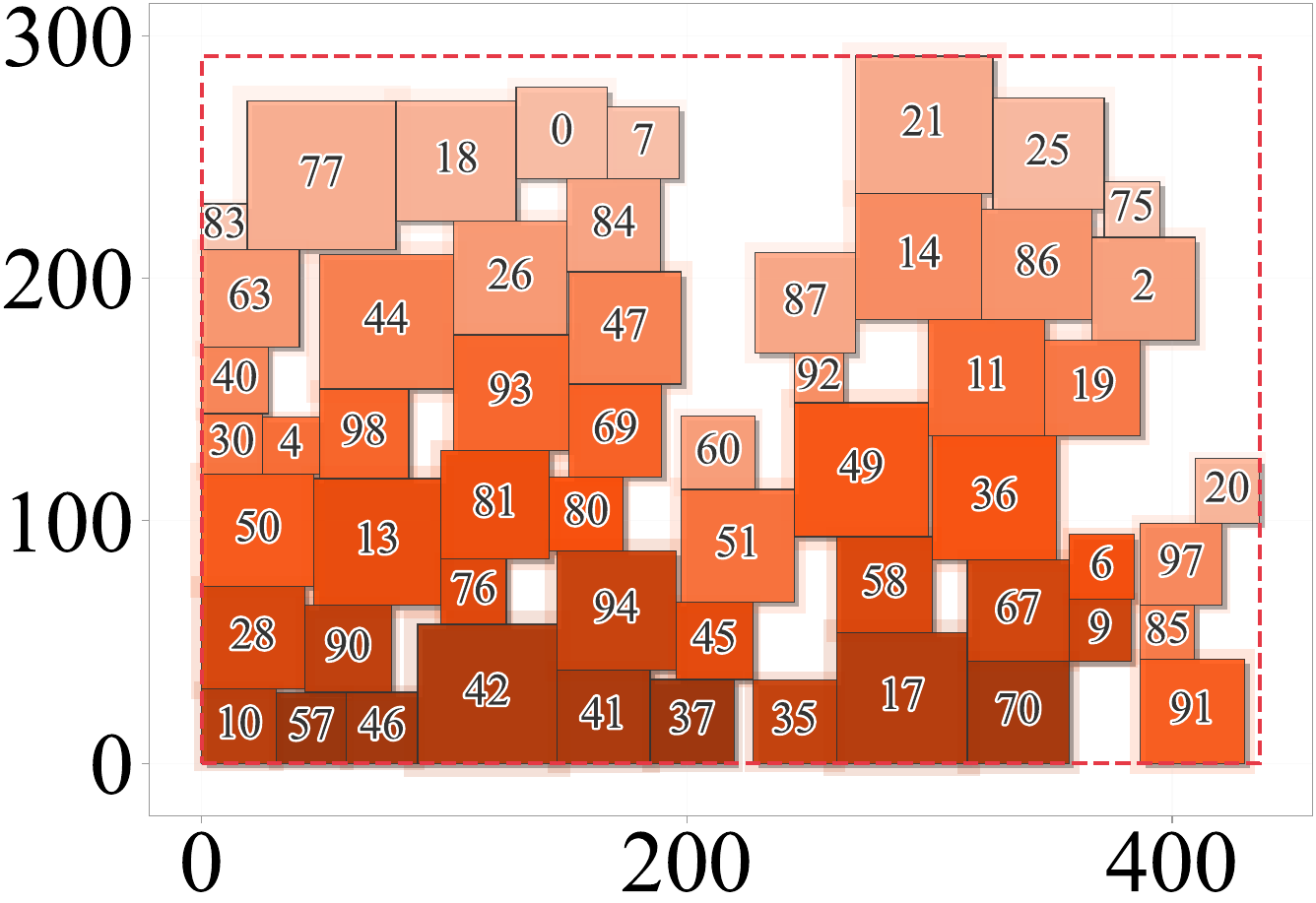}
        \captionof{figure}{Bottom die projection of the legalized 3D floorplan (aspect ratio 4:3), complementary to the top die (\Cref{fig:top_die_floorplan}).}
        \label{fig:bottom_die_floorplan}
    \end{minipage}
    \hfill
    \begin{minipage}{0.33\textwidth}
        \includegraphics[width=0.98\linewidth]{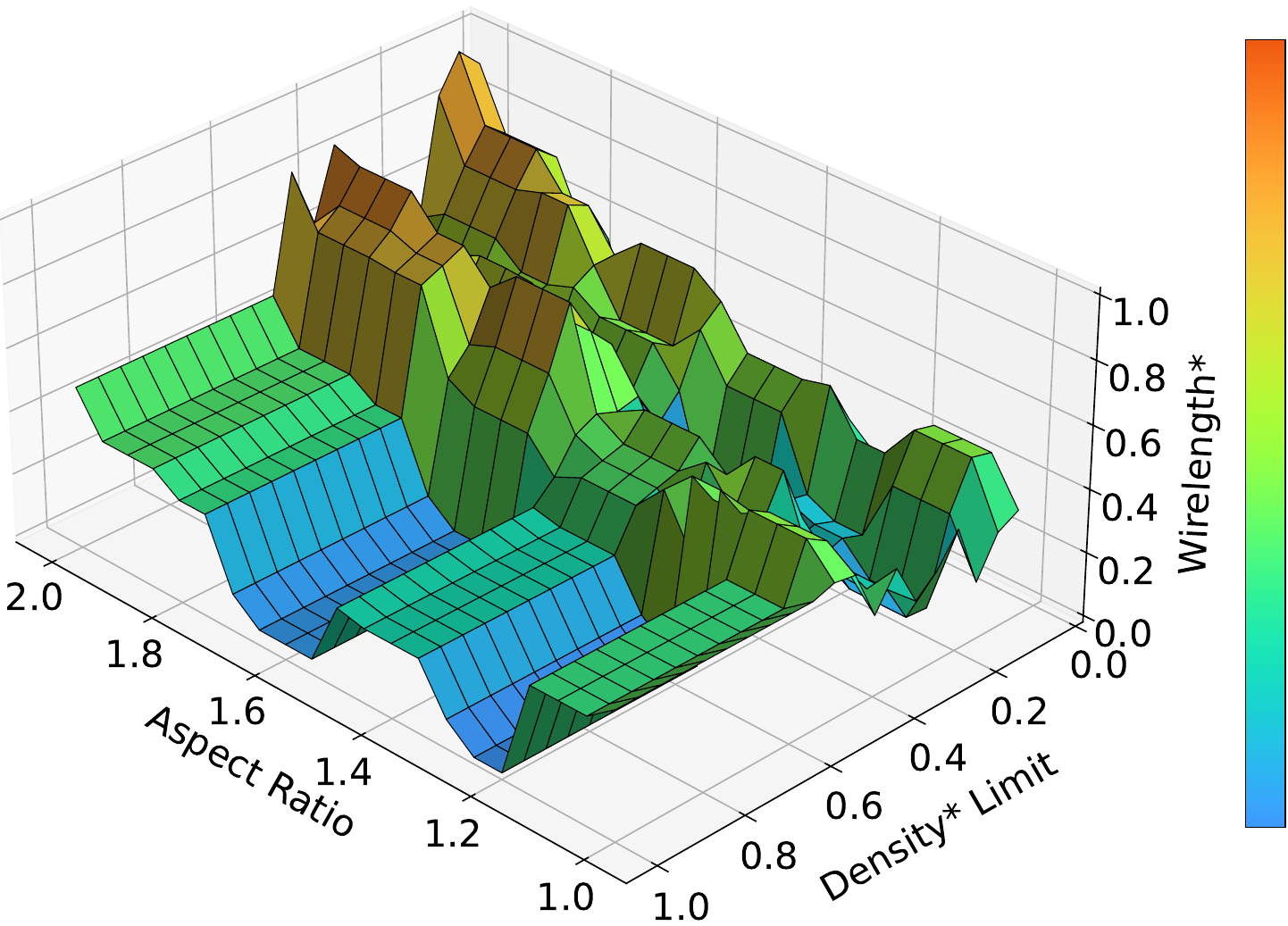}
        \captionof{figure}{Normalized wirelength surface versus per-die outline aspect ratio and inter-die density limit $\rho_{\text{max}}$ (\Cref{eq:constraint_bonding_density});
        }
        \label{fig:wirelength_surface}
    \end{minipage}
\end{figure*}
\input{tables/2-cmp_wirelength_all}
As illustrated in
\Cref{fig:top_die_floorplan,fig:bottom_die_floorplan}, the legalized
layouts satisfy the fixed-outline constraints (red dashed boxes) on both
dies while maintaining good inter-die connectivity. The complementary
block distributions across the two dies balance area usage and support
wirelength optimization, with consistent block identifiers enabling
cross-die referencing and vertical connection analysis. This final stage
bridges the gap between the continuous 3D embedding produced by the SDP
and DP stages and a 3D floorplan with soft
blocks that can be passed to detailed routing tools. Together, the three-stage
pipeline achieves a balance between global optimization (via SDP), discrete
assignment refinement (via DP), and local layout quality improvement (via
L-BFGS-B), enabling effective exploration of the 3D floorplanning design space.

%% file: algorithms/optimal_subsequence.tex
\begin{algorithm}[t]
    \caption{\textsc{Assignment via Dynamic Programming}}
    \label{alg:dp-split}
    \begin{algorithmic}[1]
        \Require $z[1\ldots n],\; Ar[1\ldots n]$
        \Ensure  assignment array $\mathbf{grp}[1\ldots n]\in\{0,1\}$
        \State $\textit{idx} \gets \operatorname*{argsort}(z)$       \label{alg:dp_sort_idx_via_z}   \Comment ascending order of $z$
        \State $Ar \gets Ar[\textit{idx}]$                       \label{alg:dp_reorder_areas}         \Comment areas reordered by $z$
        \State $P[0] \gets 0$
        \For{$k = 1$ \textbf{to} $n$}
        \State $P[k] \gets P[k-1] + Ar[k-1]$                        \Comment prefix sum
        \EndFor
        \State $T \gets \bigl(\sum_{i=1}^{n} Ar[i]\bigr)\times w / 2$  \Comment see Eq.~\eqref{eq:target_area}
        \State $best\_\delta \gets +\infty;\; best\_s \gets 0;\; best\_e \gets 0$
        \For{$e = 0$ \textbf{to} $n-1$}                                 \Comment \textbf{DP:} continuous-interval search
        \State $acc \gets 0$
        \For{$s = e$ \textbf{downto} $0$}
        \State $acc \gets acc + Ar[s]$
        \State $\delta \gets |acc - T|$
        \If{$\delta < best\_\delta$}
        \State $best\_\delta \gets \delta;\; best\_s \gets s;\; best\_e \gets e$
        \EndIf
        \EndFor
        \EndFor
        \State $\mathbf{grp}[1\ldots n] \gets 1$                        \Comment initialise all blocks to top die
        \For{$k = best\_s$ \textbf{to} $best\_e$}
        \State $\mathbf{grp}[\textit{idx}[k]] \gets 0$              \Comment assign optimal $z$-interval to bottom die
        \EndFor
        \State \Return $\mathbf{grp}$
    \end{algorithmic}
\end{algorithm}

%% file: algorithms/lbfgs.tex
\begin{algorithm}[t]
  \caption{2D Refinement via L-BFGS-B}
  \label{alg:lbfgs-refine}
  \begin{algorithmic}[1]
    \Require  Coordinates $\mathbf{X}_0$ from SDP in~\Cref{subsec:SDP_optimization}
    \Ensure   Refined coordinates $\mathbf{X}^{\ast}$

    \State  $\texttt{obj} \leftarrow \textsc{UpdateObjective}()$; \Comment{using~\Cref{eq:lbfgs_obj}}
    \State  $\mathbf{X} \gets \mathbf{X}_0$; \Comment{initialization}
    \While{convergence criteria unmet}
    \State  $g \gets \textsc{UpdateGradient()}$; \Comment{using~\Cref{eq:comp_grad}}
    \State  $\mathbf{p} \gets \textsc{UpdateDirection()}$; \Comment{via~\Cref{eq:two_loop}}
    \State  $\eta \gets \textsc{UpdateStep()}$; \Comment{using~\Cref{eq:armijo_condition}}
    \State  $\mathbf{X} \gets \mathbf{X} + \eta\,\mathbf{p}$; \Comment{position update}
    \State  \textsc{UpdateHistory()}; \Comment{using~\Cref{eq:lbfgs_s_y_pair}}
    \If{\textsc{UpdateTrigger}($\mathbf{X}$)}
    \State \textsc{UpdateParameters()}; %
    \EndIf
    \EndWhile
  \end{algorithmic}
\end{algorithm}

%% file: tables/2-cmp_wirelength_all.tex
\begin{table*}[t!]
  \centering
  \footnotesize
  \setlength{\tabcolsep}{2.6pt}
  \begin{threeparttable}
  \caption{Experimental results on GSRC benchmarks~\cite{GSRC-bench}, reporting $\mathcal{W}(\cdot)$ and runtime (seconds). Best values are in bold.}
  \label{tab:comparison_wirelength_runtime}
  \begin{tabular}{c|rr|rr|rr|rr|rr|rr|rr|rr}
    \toprule
    \multirow{2}{*}{Testcase}
         & \multicolumn{2}{c|}{FM+AR}
         & \multicolumn{2}{c|}{FM+Li}
         & \multicolumn{2}{c|}{Spectral+AR}
         & \multicolumn{2}{c|}{Spectral+Li}
         & \multicolumn{2}{c|}{GSP}
         & \multicolumn{2}{c|}{TA3D}
         & \multicolumn{2}{c|}{\cite{2026ASPDAC_Great3D}}
         & \multicolumn{2}{c}{Ours}                                                                                                                                                                                                                                                                                                                                                                                                                                                                                                                                                                      \\
    \cmidrule(lr){2-3}
    \cmidrule(lr){4-5}
    \cmidrule(lr){6-7}
    \cmidrule(lr){8-9}
    \cmidrule(lr){10-11}
    \cmidrule(lr){12-13}
    \cmidrule(lr){14-15}
    \cmidrule(lr){16-17}
         & \multicolumn{1}{c}{WL}                                & \multicolumn{1}{c|}{Time} & \multicolumn{1}{c}{WL} & \multicolumn{1}{c|}{Time} & \multicolumn{1}{c}{WL} & \multicolumn{1}{c|}{Time} & \multicolumn{1}{c}{WL} & \multicolumn{1}{c|}{Time} & \multicolumn{1}{c}{WL} & \multicolumn{1}{c|}{Time} & \multicolumn{1}{c}{WL} & \multicolumn{1}{c|}{Time} & \multicolumn{1}{c}{WL} & \multicolumn{1}{c|}{Time} & \multicolumn{1}{c}{WL} & \multicolumn{1}{c}{Time} \\

    \midrule
    n10  & 43223                                                          & 6.72                               & 48881                           & 2.16                               & 24571                           & 5.04                               & 28020                           & 2.08                               & 36444                           & 7.62                               & 41166                           & 41.10                              & 23415                           & 17.12                              & \textbf{16772}                  & \textbf{3.56}                     \\
    n30  & 84456                                                          & 13.84                              & 88703                           & 5.68                               & 67193                           & 15.68                              & 72048                           & 10.00                              & 100818                          & 41.86                              & 79429                           & 55.76                              & 56542                           & 25.20                              & \textbf{40712}                  & \textbf{12.77}                    \\
    n50  & 217731                                                         & 21.44                              & 219508                          & 18.32                              & 144133                          & 32.56                              & 147835                          & 18.82                              & 174368                          & 109.74                             & 203847                          & 159.06                             & 119495                          & 42.88                              & \textbf{71819}                  & \textbf{22.42}                    \\
    n100 & 402324                                                         & 37.04                              & 388336                          & 102.16                             & 268341                          & 45.36                              & 259307                          & 107.84                             & 385398                          & 351.90                             & 373565                          & 161.94                             & 166792                          & 91.92                              & \textbf{111715}                 & \textbf{59.42}                    \\
    n200 & 880486                                                         & 130.16                             & 852036                          & 1074.32                            & 446443                          & 115.04                             & 408192                          & 1105.04                            & 397594                          & 1360.86                            & 832867                          & 772.84                             & 345099                          & 209.68                             & \textbf{266415}                 & 289.05                            \\
    \midrule
    Avg. ratio
         & 2.9179                                                         & 1.0003                             & 2.9648                          & 1.4609                             & 1.8400                          & 1.0515                             & 1.8704                          & 1.5689                             & 2.4039                          & 4.1887                             & 2.7428                          & 5.6810                             & 1.4474                          & 2.1935                             & \textbf{1.0000}                 & \textbf{1.0000}                   \\
    \bottomrule
  \end{tabular}
  \begin{tablenotes}
    \footnotesize
    \item[*] WL: Wirelength ($\mu$m), Time: Runtime (seconds). Avg. ratio is computed as the arithmetic mean of per-case ratios normalized to ours (=1.0); lower is better. Best results achieved by our method are highlighted in \textbf{bold}.
  \end{tablenotes}
  \end{threeparttable}
\end{table*}

%% file: doc/4-experiment.tex
\section{Experimental Results}
\label{sec:exp}
\subsection{Experimental Setup}
\label{subsec:exp_setup}
All experiments are conducted on a Linux workstation running Ubuntu 22.04 LTS, equipped with dual Intel Xeon Gold 6426Y processors and 256~GB RAM. Our framework is implemented in Python~3.11.
We evaluate \textsc{Great3D} (this work) on the GSRC benchmark suite~\cite{GSRC-bench} and ICCAD 2024 ATPlace benchmark suite~\cite{2024ICCAD-ATPlace2.5D}, enabling comprehensive assessment of scalability and solution quality. For GSRC~\cite{GSRC-bench}, we directly use the provided soft block areas and netlists to form 3D floorplanning instances under the specified outline constraints.
For ATPlace~\cite{2024ICCAD-ATPlace2.5D}, we follow the benchmark specifications and extract block areas and connectivity to construct 3D floorplanning instances under the given outlines.
\textcolor{blue}{The CPI-model parameters used in the latency-aware evaluation are listed in \Cref{tab_parameters}; they are taken from prior architecture-level studies~\cite{parameters-clapp2015quantifying,2010TVLSI-parameters_wire2latency,2023ICCAD-Zhuang-3d-bonding} and are identical for all evaluated methods.}
\input{tables/0_exp_paras}
We compare \textsc{Great3D} (this work) against the following seven representative baselines widely used in prior 3D floorplanning studies for fair comparison:
\begin{itemize}
  \item FM+AR, which combines the FM partitioning algorithm~\cite{1995ICCAD-FM_partition} with the AR 2D floorplanner~\cite{2008ASPDAC-Luo-AR};
  \item FM+Li, using FM partitioning and Li's 2D optimizer~\cite{2023DAC-Li};
  \item Spectral+AR, using spectral partitioning~\cite{1995DAC-Spectral_partitioning} with AR~\cite{2008ASPDAC-Luo-AR};
  \item Spectral+Li, combining spectral partitioning with Li's floorplanner~\cite{2023DAC-Li};
  \item GSP, a grouped sequence-pair-based 3D floorplanner~\cite{2011TVLSI-grouped-sequence-pair};
  \item TA3D, a thermal-aware 3D floorplanner using smoothed HPWL and FM-style optimization~\cite{2021TVLSI-thermal-aware-3dfloorplan-tsv};
  \item \textsc{Great3D} (median-cut), our earlier native-3D framework using median-cut for die assignment~\cite{2026ASPDAC_Great3D}.
\end{itemize}
\noindent The first six baselines represent state-of-the-art 3D floorplanning approaches from the literature, while the last baseline enables direct assessment of the DP-based refinement's contribution. \textcolor{blue}{The classical sequence-pair-based 3D floorplanners~\cite{2010Integration-Partition-Sequence-pair} encode block ordering via (grouped) sequence pairs explored with simulated-annealing-style metaheuristics; GSP~\cite{2011TVLSI-grouped-sequence-pair} and TA3D~\cite{2021TVLSI-thermal-aware-3dfloorplan-tsv} serve as representatives of this family.} We report wirelength (HPWL), runtime, and CPI under the same benchmark settings and outline constraints.
These comparisons evaluate the algorithmic benefit of the proposed partitioning-free floorplanning strategy under a common block-level benchmark abstraction.
\subsection{Performance Comparison with State-of-the-Art}
\label{subsec:cmp_results}
\Cref{tab:comparison_wirelength_runtime} compares \textsc{Great3D} (this work) with seven baselines across five GSRC testcases, reporting both wirelength and runtime.
\textsc{Great3D} (this work) consistently achieves the best wirelength among all methods, demonstrating the effectiveness of the unified native-3D optimization combined with DP-based die-assignment refinement.

\mysection{Comparison with partitioning-based methods (Baselines 1--4):}
Traditional partition-first approaches (FM+AR, FM+Li, Spectral+AR, Spectral+Li) commit to die assignments before floorplanning, which can limit their optimization potential. While these methods may run slightly faster, their early-stage partitioning leads to significantly worse wirelength (1.84--2.96$\times$ worse on average). This validates our key insight that joint optimization of die assignment and block placement is essential for high-quality 3D floorplanning.
\mysection{Comparison with 3D-native methods (Baselines 5--6):}
3D-native approaches (GSP, TA3D) that extend 2D representations face exponential search-space explosion along the $z$-axis, resulting in longer runtimes (4.19--5.68$\times$ on average) without delivering better wirelength (2.40--2.74$\times$ worse). In contrast, \textsc{Great3D} (this work)'s SDP-based global placement provides a tractable continuous relaxation that avoids combinatorial complexity while capturing 3D spatial relationships.

\mysection{Comparison with \textsc{Great3D} (median-cut) (Baseline 7):}
The DP-based refinement introduced in \textsc{Great3D} (this work) improves wirelength by 1.45$\times$ and achieves a 2.19$\times$ speedup on average over the median-cut baseline~\cite{2026ASPDAC_Great3D} on GSRC benchmarks (\Cref{tab:gsrc_ablation}).
This substantial improvement demonstrates the importance of explicit die-area balance and connectivity-aware splitting, addressing the two key limitations of the simple median-cut heuristic identified in~\Cref{subsec:overview}.

Overall, \textsc{Great3D} (this work) maintains competitive runtime while consistently achieving the best wirelength across all GSRC testcases, demonstrating strong Pareto efficiency.

\mysection{\textcolor{blue}{Inter-die connectivity and die-area balance:}}
\textcolor{blue}{At the block level, the inter-die wirelength quantifies the amount of face-to-face (F2F) connectivity, and is best read jointly with total wirelength and die-area balance, defined as $\min(A_0,A_1)/\max(A_0,A_1)$, where $A_0$ and $A_1$ are the total block areas placed on the bottom and top die, respectively ($1.0$ = perfectly balanced). \Cref{tab:interdie_f2f} reports these metrics across the benchmark suite, with ``Inter-die \%'' giving the inter-die wirelength as a fraction of total wirelength. \textsc{Great3D} attains the lowest total wirelength together with the best die-area balance ($0.962$, i.e., nearly equal die utilization) while keeping its inter-die connectivity at a balanced level. The spectral-based methods reach a lower inter-die wirelength only by confining strongly-connected blocks to a single die, which inflates their total wirelength to roughly $1.9\times$ that of \textsc{Great3D} ($154.1$ vs.\ $80.4$). This confirms that the inter-die connectivity of \textsc{Great3D} reflects a genuinely balanced partitioning-free 3D assignment.}

\input{tables/6-2.0vs1.0}
We then use \Cref{eq:3dflp_hybrid-bonding-latency} with the parameters in \Cref{tab_parameters} to compute CPI on the GSRC benchmarks (\Cref{fig:cpi_comparison}, \textcolor{blue}{lower is better}). Because \textsc{Great3D} co-optimizes die assignment and block placement, it shortens the critical cross-die paths captured by $L$ while simultaneously minimizing $\text{wirelength}_{ppi}$, which (as noted in \Cref{sec:prelim}) lowers both wirelength and the latency-induced CPI component. \textcolor{blue}{Across all GSRC cases, the full \textsc{Great3D} (``Ours with DP'') attains the lowest CPI among the compared methods, and further reduces CPI by $9.5$--$17.8\%$ over the ablation without the DP die-assignment stage (``Ours w/o DP''), as the per-case bars in \Cref{fig:cpi_comparison} show.} This demonstrates the benefit of the DP-enhanced latency-aware native 3D flow.

\input{tables/5_atplace_bench_table}

\input{tables/7-interdie_f2f}
\mysection{Stability analysis:}
To further evaluate the robustness of our approach, we conduct a comprehensive stability analysis across 10 random seeds on the ATPlace benchmark suite (10 cases) for each method.
\Cref{fig:boxplot_cpi} presents the CPI distribution as a boxplot, visualizing the median, interquartile range, and outliers across methods.
\Cref{fig:bar_hpwl} shows the average HPWL with standard deviation for each method, comparing both solution quality and variability under different random seeds.

On the full ATPlace suite, \textsc{Great3D} attains the best aggregate normalized ratio ($1.00$) across all metrics in \Cref{tab:full_comparison}; since each baseline implicitly optimizes a different objective, no method dominates every individual case, yet \textsc{Great3D} delivers the most balanced overall trade-off across wirelength, area, and CPI.

As illustrated in \Cref{fig:top_die_floorplan,fig:bottom_die_floorplan}, modules are placed to optimize inter-die connectivity while respecting die boundaries (red dashed boxes), and the complementary placement across dies balances area usage. \Cref{fig:wirelength_surface} shows the normalized wirelength surface under varying \textcolor{blue}{per-die outline aspect ratio (width/height of each die outline, restricted to $[1.0,2.0]$ to avoid symmetry redundancy) and density limit $\rho_{\text{max}}$ of \Cref{eq:constraint_bonding_density}}. Reducing $\rho_{\text{max}}$ at a fixed aspect ratio limits cross-die placement flexibility and degrades wirelength, whereas the aspect ratio affects wirelength non-monotonically due to the discrete nature of block configurations.

\input{tables/8-backend_grt}

\mysection{\textcolor{blue}{Backend global-route validation:}}
\textcolor{blue}{To evaluate the generated floorplans in a real 3D backend implementation flow, we feed each floorplan into the Open3DBench 3D-IC backend framework~\cite{shi2025open3dbench}, which drives the OpenROAD routing engine~\cite{ajayi2019toward} over a two-tier NanGate45 stack. We evaluate the \texttt{ariane133} design from OpenROAD-flow-scripts~\cite{openroadflow}: each method patches the shared DEF with its own macro locations and tier assignment, after which OpenROAD runs \texttt{global\_route} and reports global-route wirelength and congestion overflow (\Cref{tab:backend_grt}). \textsc{Great3D} carries its block-level advantage through to the downstream flow, attaining both the lowest global-route wirelength and by far the lowest congestion overflow ($91.4\%$ lower than \textsc{Fm}-based and $74.5\%$ lower than \textsc{Ta3d}). This congestion-overflow margin, the largest gap across all backend metrics, reflects markedly fewer routing-resource conflicts on the bonded two-tier stack, showing that on \texttt{ariane133}, our method produces the most routable layout among all compared methods and a high-quality initial solution for subsequent 3D IC implementation.}

%% file: tables/0_exp_paras.tex
\begin{table}[t!]
    \centering
    \footnotesize
    \setlength{\tabcolsep}{4pt}
    \caption{Parameter Settings on CPI\tnote{*}.}
    \label{tab_parameters}
    \begin{tabular}{c|r|c||c|l|c}
      \toprule
      \textbf{Parameter} & \multicolumn{1}{c|}{\textbf{Value}} & \textbf{Ref.}                          & \textbf{Parameter} & \multicolumn{1}{c|}{\textbf{Value}} & \textbf{Ref.}                            \\
      \midrule
      $\alpha_1$         & $5\times10^{-4}$                    & \cite{parameters-clapp2015quantifying} & $\alpha_2$         & $1.8\times10^{-1}$                  & \cite{parameters-clapp2015quantifying}   \\
      $\beta_1$          & $1.1\times10^{-1}$                  & \cite{parameters-clapp2015quantifying} & $MP$               & $2.4\times10^{-1}$                  & \cite{parameters-clapp2015quantifying}   \\
      $CPI_{sta}$        & \textcolor{blue}{$1.532\times10^{-1}$} & \cite{2023ICCAD-Zhuang-3d-bonding}  & $\beta_2$          & $9.8 \times 10^1$                   & \cite{2010TVLSI-parameters_wire2latency} \\
      \bottomrule
    \end{tabular}
\end{table}

%% file: tables/6-2.0vs1.0.tex
\begin{table}[t!]
  \centering
  \color{blue}%
  \footnotesize
  \setlength{\tabcolsep}{4pt}
  \caption{\textcolor{blue}{DP die-assignment ablation on GSRC: ``w/o DP'' $=$ SDP\,+\,2D refinement, ``with DP'' $=$ full \textsc{Great3D}; ``Ratio'' is normalized to ``with DP'' (Ours).}}
  \label{tab:gsrc_ablation}
  \resizebox{\columnwidth}{!}{%
  \color{blue}%
  \begin{tabular}{ll|rrrrr|r}
    \toprule
    \textbf{Metric} & \textbf{Config} & \textbf{n10} & \textbf{n30} & \textbf{n50} & \textbf{n100} & \textbf{n200} & \textbf{Ratio} \\
    \midrule
    \multirow{2}{*}{WL ($10^3\mu$m)}
       & w/o DP~\cite{2026ASPDAC_Great3D}  & 23.4          & 56.5          & 119.5         & 166.8          & 345.1          & 1.45          \\
       & with DP (Ours)                    & \textbf{16.8} & \textbf{40.7} & \textbf{71.8} & \textbf{111.7} & \textbf{266.4} & \textbf{1.00} \\
    \midrule
    \multirow{2}{*}{Time (s)}
       & w/o DP~\cite{2026ASPDAC_Great3D}  & 17.12         & 25.20         & 42.88         & 91.92          & 209.68         & 2.19          \\
       & with DP (Ours)                    & \textbf{3.56} & \textbf{12.77}& \textbf{22.42}& \textbf{59.42} & 289.05         & \textbf{1.00} \\
    \bottomrule
  \end{tabular}%
  }
\end{table}

%% file: tables/5_atplace_bench_table.tex
\begin{table*}[t]
  \centering
  \footnotesize
  \setlength{\tabcolsep}{3pt}
  \caption{Performance comparison of 3D floorplanning methods on the ICCAD 2024 ATPlace benchmark suite~\cite{2024ICCAD-ATPlace2.5D}.}
  \label{tab:full_comparison}
  \begin{threeparttable}
  \begin{tabular}{c|ccc|ccc|ccc|ccc|ccc|ccc|ccc}
    \toprule
    \multirow{2}{*}{Case} & \multicolumn{3}{c|}{FM+AR} & \multicolumn{3}{c|}{FM+Li} & \multicolumn{3}{c|}{Spec+AR} & \multicolumn{3}{c|}{Spec+Li} & \multicolumn{3}{c|}{GSP} & \multicolumn{3}{c|}{TA3D} & \multicolumn{3}{c}{\textsc{Great3D} (this work)} \\
    \cmidrule{2-22}
                          & WL & Area & CPI
                          & WL & Area & CPI
                          & WL & Area & CPI
                          & WL & Area & CPI
                          & WL & Area & CPI
                          & WL & Area & CPI
                          & WL & Area & CPI \\
    \midrule
    1  & 53.6 & 739 & 8.62  & 55.8 & 769 & 8.96  & 70.5 & 820 & 11.28 & 70.5 & 820 & 11.28 & \textbf{45.4} & 743 & \textbf{7.32} & 50.6 & \textbf{693} & 8.14  & 54.6 & 729 & 8.77 \\
    2  & 62.1 & 861 & 8.98  & 62.2 & 897 & 8.99  & 60.8 & 871 & 8.79  & 58.7 & 907 & 8.49  & \textbf{48.8} & 835 & \textbf{7.09} & 68.6 & 880 & 9.90  & 55.2 & \textbf{708} & 8.00 \\
    3  & 106.0& 408 & 6.43  & 106.0& 408 & 6.43  & 119.7& 372 & 7.24  & 121.0& 366 & 7.32  & 81.5 & 394 & 4.97  & 100.5& 403 & 6.10  & \textbf{28.7} & \textbf{357} & \textbf{1.85} \\
    4  & 163.9& 836 & 11.24 & 162.8& 836 & 11.17 & 202.1& 842 & 13.82 & 202.9& 844 & 13.88 & 101.4& 728 & 7.02  & 120.1& 717 & 8.28  & \textbf{54.7} & \textbf{693} & \textbf{3.85} \\
    5  & 197.4& 1863& 14.18 & 193.6& 1829& 13.90 & 257.1& 1844& 18.41 & 256.5& 1844& 18.37 & 153.7& 1696& 11.07 & 192.4& \textbf{1644} & 13.82 & \textbf{150.2} & 1650 & \textbf{10.82} \\
    6  & 171.7& 1048& 15.40 & 167.7& 1041& 15.04 & 211.0& 1126& 18.88 & 203.4& 1137& 18.21 & 109.9& 1061& 9.91  & \textbf{106.5} & 976 & \textbf{9.61} & 120.8& \textbf{955} & 10.88 \\
    7  & 39.2 & 343 & 7.11  & 39.5 & 314 & 7.17  & 32.5 & 281 & 5.92  & 44.2 & 284 & 8.00  & 28.3 & 314 & 5.19  & 28.9 & 273 & 5.29  & \textbf{25.4} & \textbf{266} & \textbf{4.66} \\
    8  & 27.8 & 220 & 4.87  & 35.5 & 227 & 6.18  & 41.9 & 225 & 7.26  & 49.6 & 227 & 8.56  & 28.3 & 245 & 4.96  & 26.9 & \textbf{199} & 4.72  & \textbf{23.0} & 218 & \textbf{4.05} \\
    9  & 258.7& 1757& 17.05 & 293.1& 1912& 19.30 & 374.9& 1867& 24.63 & 372.7& 1822& 24.49 & 197.4& 1767& 13.04 & \textbf{186.0} & 1498& \textbf{12.30} & 190.0& \textbf{1465} & 12.56 \\
    10 & 157.9& 1090& 15.11 & 161.3& 1098& 15.43 & 170.7& 1023& 16.32 & 176.7& 1003& 16.89 & 114.7& 1078& 11.01 & \textbf{94.9} & \textbf{907} & \textbf{9.14} & 101.1& 931 & 9.72 \\
    \midrule
    Ratio & 1.72 & 1.15 & 1.69
          & 1.77 & 1.16 & 1.74
          & 2.05 & 1.14 & 2.00
          & 2.13 & 1.14 & 2.08
          & 1.29 & 1.12 & 1.27
          & 1.43 & 1.03 & 1.40
          & \textbf{1.00} & \textbf{1.00} & \textbf{1.00} \\
    \bottomrule
  \end{tabular}
  \begin{tablenotes}[flushleft]
    \footnotesize
    \item[*] WL (wirelength, M$\mu$m), Area (M$\mu$m$^2$, max of top/bottom die), and CPI are reported as the mean over 10 independent runs with different random seeds. The last row shows the ratio of each method's mean to ours; lower is better.
    \end{tablenotes}
\end{threeparttable}
\end{table*}

%% file: tables/7-interdie_f2f.tex
\begin{table}[t!]
  \centering
  \color{blue}%
  \footnotesize
  \setlength{\tabcolsep}{4pt}
  \caption{\textcolor{blue}{Block-level F2F connectivity proxy, averaged over all benchmark cases (HPWL in $10^{6}\,\mu$m). Best values are in bold.}}
  \label{tab:interdie_f2f}
  \begin{tabular}{l|rrrr|r}
    \toprule
    \textbf{Metric}
      & \textsc{Gsp} & \textsc{Ta3d} & \textsc{Fm} & Spec. & \textbf{\textsc{Great3D}} \\
      & \cite{2011TVLSI-grouped-sequence-pair} & \cite{2021TVLSI-thermal-aware-3dfloorplan-tsv} & \cite{1995ICCAD-FM_partition} & \cite{1995DAC-Spectral_partitioning} & \textbf{(ours)} \\
    \midrule
    Total WL ($10^{6}\mu$m)     & 90.9  & 97.5  & 123.8 & 154.1 & \textbf{80.4}  \\
    Inter-die WL ($10^{6}\mu$m) & 34.5  & 49.6  & 47.0  & 19.2  & 39.4           \\
    Inter-die \%                & 38    & 51    & 38    & 12    & 49             \\
    Area bal.                   & 0.494 & 0.835 & 0.621 & 0.875 & \textbf{0.962} \\
    \bottomrule
  \end{tabular}
\end{table}

%% file: tables/8-backend_grt.tex
\begin{table}[t!]
  \centering
  \color{blue}%
  \footnotesize
  \setlength{\tabcolsep}{3pt}
  \caption{\textcolor{blue}{Backend global-route results on the \texttt{ariane133} design~\cite{openroadflow} via the Open3DBench framework~\cite{shi2025open3dbench} driving the OpenROAD router~\cite{ajayi2019toward}; ``norm.'' is each baseline normalized to \textsc{Great3D} (lower is better).}}
  \label{tab:backend_grt}
  \resizebox{\columnwidth}{!}{%
  \color{blue}%
  \begin{tabular}{l|rr|rr|rr|r}
    \toprule
    \multirow{2}{*}{\textbf{Metric}}
      & \multicolumn{2}{c|}{\textsc{Fm}~\cite{1995ICCAD-FM_partition}}
      & \multicolumn{2}{c|}{Spec.~\cite{1995DAC-Spectral_partitioning}}
      & \multicolumn{2}{c|}{\textsc{Ta3d}~\cite{2021TVLSI-thermal-aware-3dfloorplan-tsv}}
      & \textbf{\textsc{Great3D}} \\
      & val & norm. & val & norm. & val & norm. & \textbf{(ours)} \\
    \midrule
    GRT WL ($10^{6}\mu$m) & 7.89    & 1.09  & 7.87    & 1.09  & 7.30    & 1.01 & \textbf{7.21} \\
    Overflow              & 4{,}651 & 11.66 & 3{,}995 & 10.01 & 1{,}562 & 3.92 & \textbf{399}  \\
    \bottomrule
  \end{tabular}%
  }
\end{table}

%% file: doc/5-conclusion.tex
\begin{figure}[t]
    \centering
    \subfloat[CPI distribution]{%
        \includegraphics[width=0.49\linewidth]{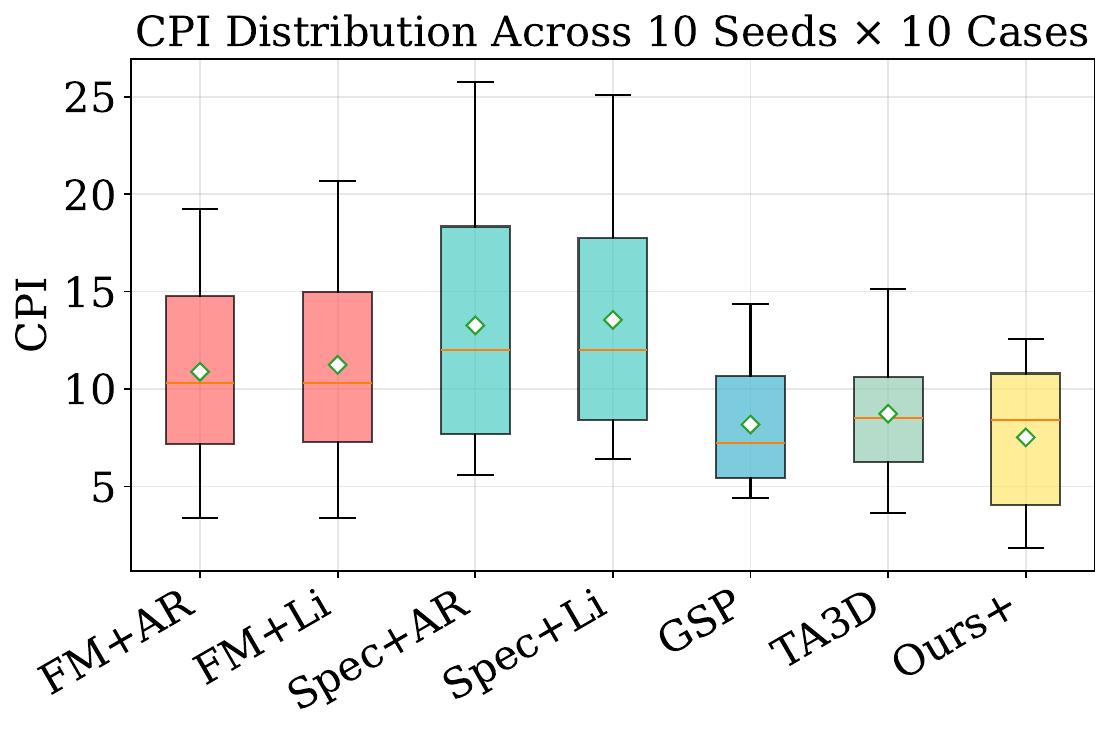}%
        \label{fig:boxplot_cpi}%
    }
    \hfill
    \subfloat[Average HPWL]{%
        \includegraphics[width=0.49\linewidth]{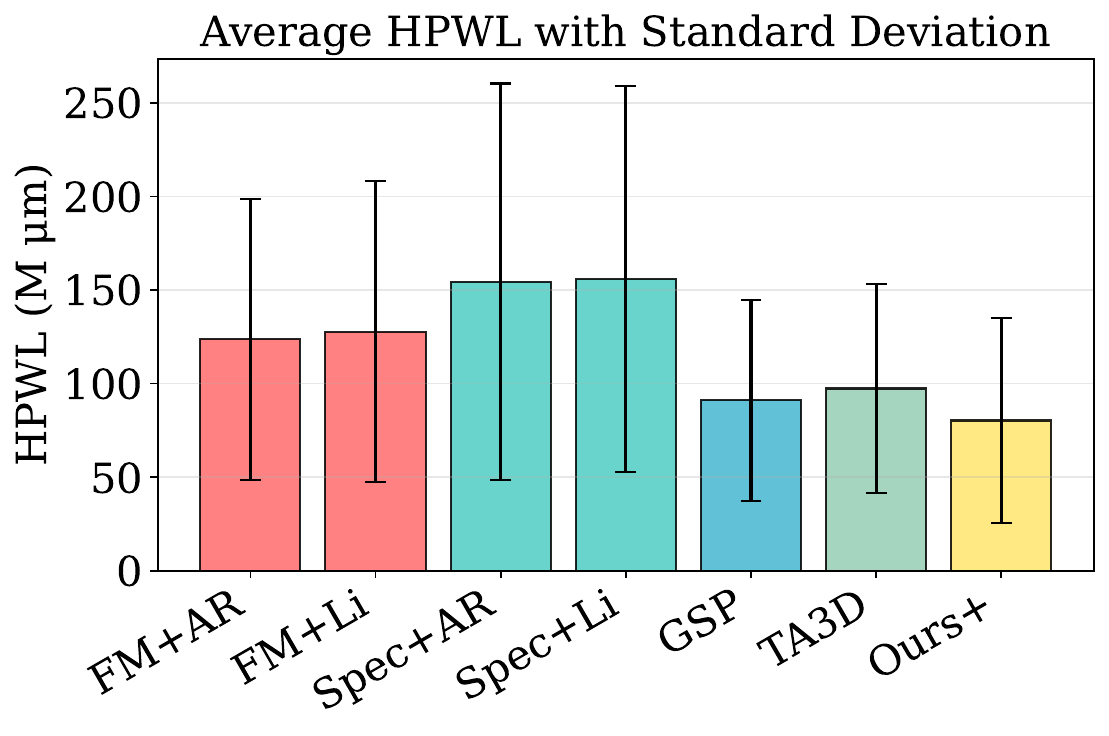}%
        \label{fig:bar_hpwl}%
    }
    \caption{Multi-seed stability across the ATPlace benchmark suite (10 cases, 10 random seeds).
    (a)~CPI distribution as a boxplot, showing the median, interquartile range, and outliers.
    (b)~Average HPWL with standard-deviation error bars, summarizing solution quality and consistency across methods.}
    \label{fig:multiseed_stability}
\end{figure}

\section{Conclusion}
\label{sec:conclusion}

This paper presented \textsc{Great3D}, a native 3D floorplanning framework that jointly optimizes die assignment and block placement within a unified analytical formulation.
By operating directly in the 3D solution space and avoiding early netlist partitioning, the proposed method enables effective coordination between global connectivity optimization and physical layout constraints.
The integration of continuous 3D embedding with discrete die-assignment refinement enables \textsc{Great3D} to consistently achieve strong wirelength quality across the GSRC and ATPlace benchmark suites; \textcolor{blue}{on GSRC, the DP die-assignment stage further improves CPI by $9.5$--$17.8\%$ over the variant without it.}
\textcolor{blue}{This work focuses on partitioning-free, latency-aware 3D floorplanning; thermal-aware co-optimization, which can be folded into the unified objective as an additional weighted term, is an important but separate problem left for future work.}
Overall, this work establishes a robust and extensible foundation for native 3D IC floorplanning and demonstrates the practical advantages of analytical 3D optimization.

\section*{Acknowledgment}
This work was conducted in the JC STEM Lab of Intelligent Design Automation, funded by The Hong Kong Jockey Club Charities Trust. This work was supported in part by the Research Grants Council of the Hong Kong Special Administrative Region, China, under Grant Nos. CUHK14211324 and CUHK7010840, and in part by ACCESS - AI Chip Center for Emerging Smart Systems, supported by the InnoHK initiative of the Innovation and Technology Commission of the Hong Kong Special Administrative Region Government.
\textcolor{blue}{AI assistants were used for language polishing and partial code generation under the authors' supervision. All research decisions and technical contributions were made by the authors.}

%% file: doc/bio.tex
\vspace{-.2in}
\begin{IEEEbiography}
    [{\includegraphics[width=1.0in,height=1.26in,clip,keepaspectratio]{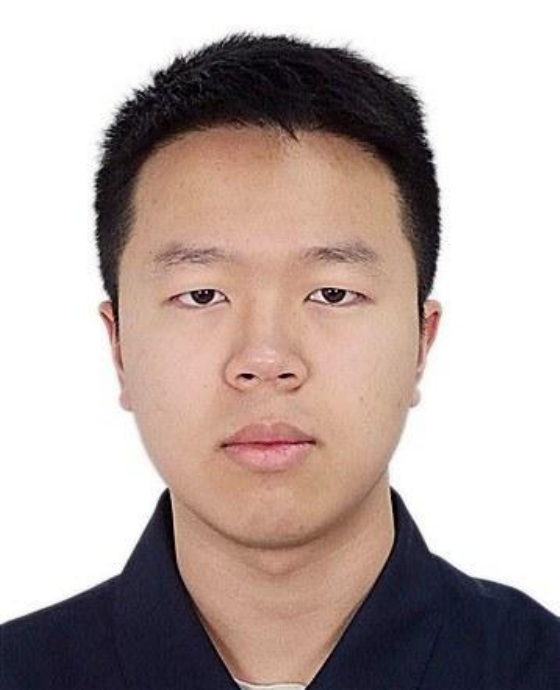}}]
    {Shuo~Ren}
    received his B.S. degree in computer science from the Huazhong University of Science and Technology, Wuhan, China, in 2024. He is currently working toward the Ph.D. degree in the Department of Computer Science and Engineering, The Chinese University of Hong Kong. His current research focuses on electronic design automation, with particular interests in 3D integrated circuit design and physical design optimization.
\end{IEEEbiography}

\vspace{-.2in}
\begin{IEEEbiography}
    [{\includegraphics[width=1.0in,height=1.26in,clip,keepaspectratio]{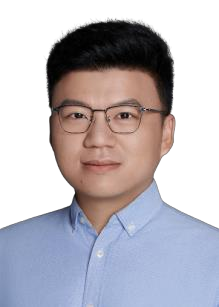}}]
    {Rongliang~Fu}
    received his Ph.D. in Computer Science and Engineering from The Chinese University of Hong Kong in March 2026, following an M.S. from the University of Chinese Academy of Sciences in June 2021 and a B.S. in Software Engineering from Northwestern Polytechnical University in June 2018. He was awarded the Hong Kong RGC JRFS 2026/27 and has authored over 40 papers across major journals (IEEE TC and IEEE TCAD) and conferences(DAC, DATE, ICCAD, etc.). His research spans electronic design automation (EDA), especially in logic optimization and EDA for superconducting electronics.
\end{IEEEbiography}

\vspace{-.2in}
\begin{IEEEbiography}
    [{\includegraphics[width=1.0in,height=1.26in,clip,keepaspectratio]{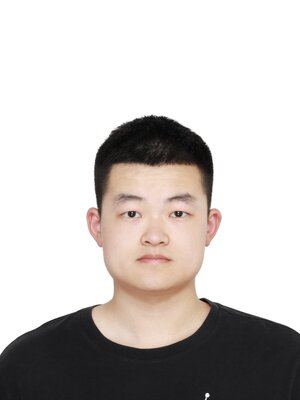}}]
    {Libo Shen}
    received his B.S. degree in communication engineering from Beijing University of Posts and Telecommunications, Beijing, China, in 2021 and his MS degree in computer technology from the University of Chinese Academy of Sciences, Beijing, China, in 2024. He is currently a Ph.D. student at the Department of Computer Science and Engineering, The Chinese University of Hong Kong. His research interests include electronic design automation and computer architecture.
\end{IEEEbiography}

\vspace{-.2in}
\begin{IEEEbiography}
    [{\includegraphics[width=1.0in,height=1.26in,clip,keepaspectratio]{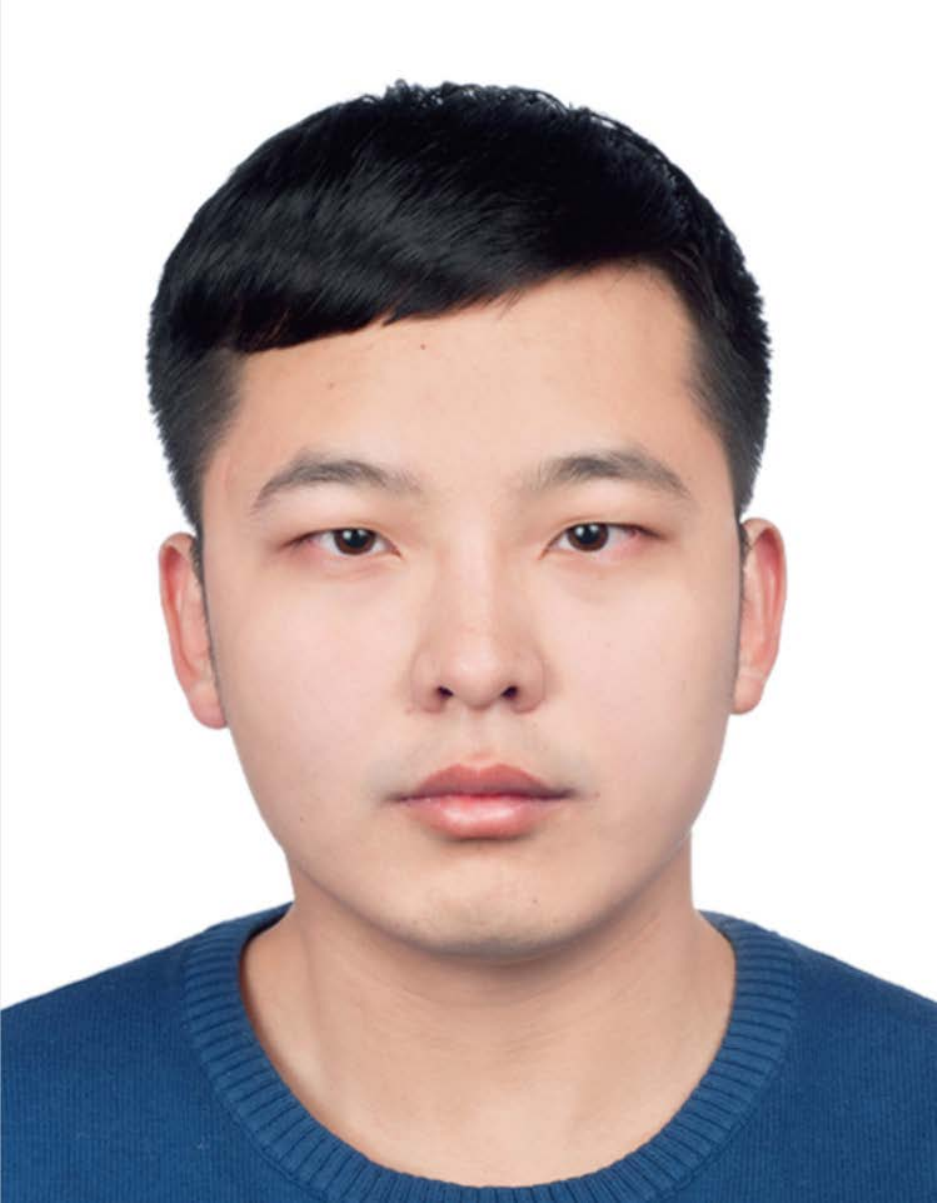}}]
    {Zhen Zhuang}
    received his M.Eng. and B.Eng. from Fuzhou University in 2021 and 2018, respectively. He obtained his Ph.D. from the Chinese University of Hong Kong in 2025 under the supervision of Prof. Tsung-Yi Ho. His current research interest is Electronic Design Automation (EDA), especially EDA for advanced packaging and 3D IC. He was a recipient of three ICCAD/ISPD contest awards.
\end{IEEEbiography}

\vspace{-.2in}
\begin{IEEEbiography}
    [{\includegraphics[width=1.0in,height=1.26in,clip,keepaspectratio]{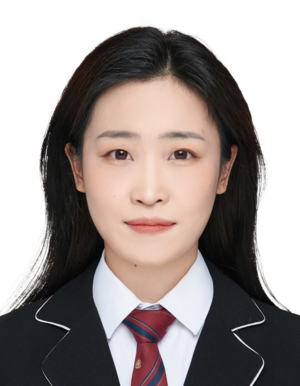}}]
    {Leilei Jin}
    is currently a postdoctoral researcher affiliated with the Department of Computer Science and Engineering at The Chinese University of Hong Kong (CUHK). She received her Ph.D. degree from Southeast University in 2024, where her doctoral work focused on foundational theories and practical methodologies in integrated circuit design automation. Her research interests include static timing analysis, crosstalk prediction, PPA optimization for Backside PDN and 3DICs under advanced process nodes.
\end{IEEEbiography}

\vspace{-.4in}
\begin{IEEEbiography}
    [{\includegraphics[width=1.0in,height=1.26in,clip,keepaspectratio]{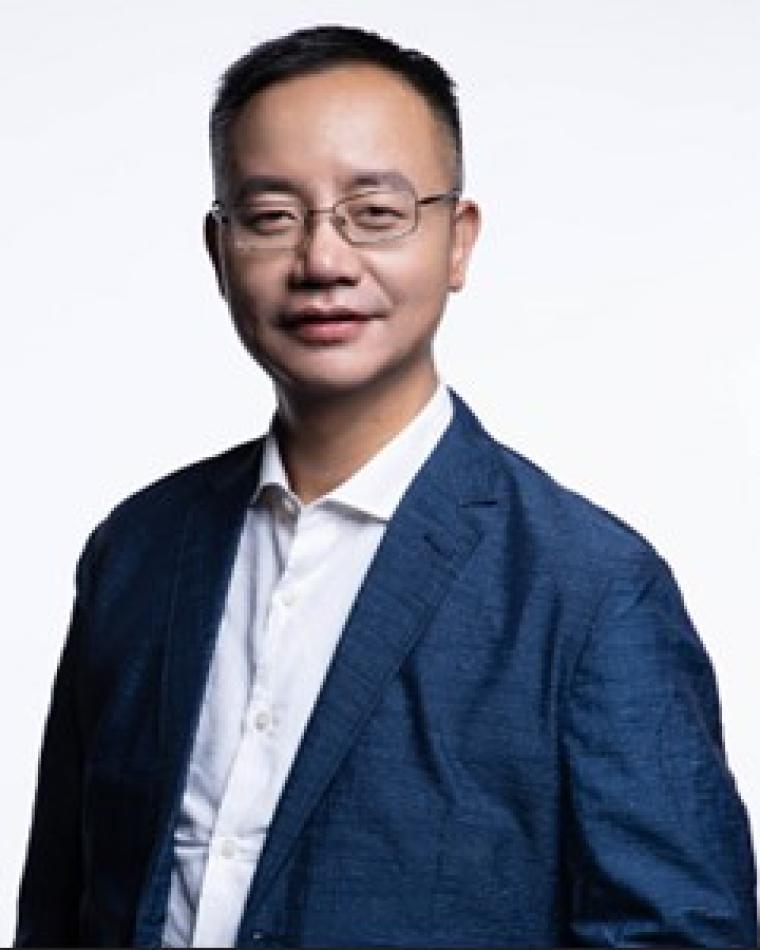}}]
    {Hao Yu}
    (Senior Member, IEEE)
    received the Ph.D. degree in electrical engineering from the University of California at Los Angeles, USA, in 2007. He is currently with the Southern University of Science and Technology, Shenzhen, China. His research interests include energy-efficient data links, sensors, and analysis. He is a Senior Member of ACM. He has about 282 peer-reviewed IEEE/ACM publications; three best paper award nominations for Design Automation Conference in 2006, International Conference on Computer-Aided Design in 2006, and International Conference on VLSI Design Automation in Asia and South Pacific Region in 2012. He is an Associate Editor of \textit{Scientific Reports} (Nature), \textit{IEEE Transactions on Biomedical Circuits and Systems}, \textit{ACM Transactions on Embedded Computing Systems}, and \textit{Microelectronics} (Elsevier); and a Technical Program Committee Member of IEEE Custom Integrated Circuits Conference, IEEE Asian Solid-State Circuits Conference, ACM-DAC, and ACM Design, Automation and Test in Europe Conference and Exhibition; and of many IEEE/ACM international journals and conferences.
\end{IEEEbiography}

\vspace{-.4in}
\begin{IEEEbiography}
    [{\includegraphics[width=1.0in,height=1.26in,clip,keepaspectratio]{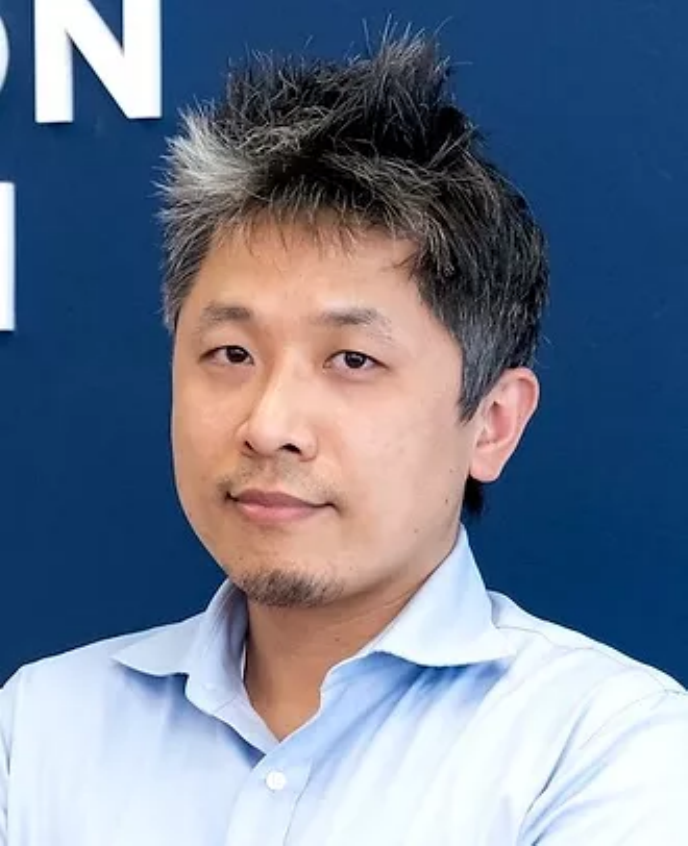}}]
    {Bei Yu}
    (M'15-SM'22)
    received the Ph.D.~degree from The University of Texas at Austin in 2014.
    He is currently a Professor in the Department of Computer Science and Engineering, The Chinese University of Hong Kong.
    He has served as TPC Chair of ACM/IEEE Workshop on Machine Learning for CAD, and in many journal editorial boards and conference committees.
    He received ten Best Paper Awards from IEEE TSM 2022, DATE 2022, ICCAD 2021 \& 2013, ASPDAC 2021 \& 2012, ICTAI 2019, Integration, the VLSI Journal in 2018, ISPD 2017, SPIE Advanced Lithography Conference 2016, and many other awards, including DAC Under-40 Innovator Award (2024), IEEE CEDA Ernest S.~Kuh Early Career Award (2022), and Hong Kong RGC Research Fellowship Scheme (RFS) Award (2024).
\end{IEEEbiography}

\vspace{-.4in}
\begin{IEEEbiography}
    [{\includegraphics[width=1.0in,height=1.26in,clip,keepaspectratio]{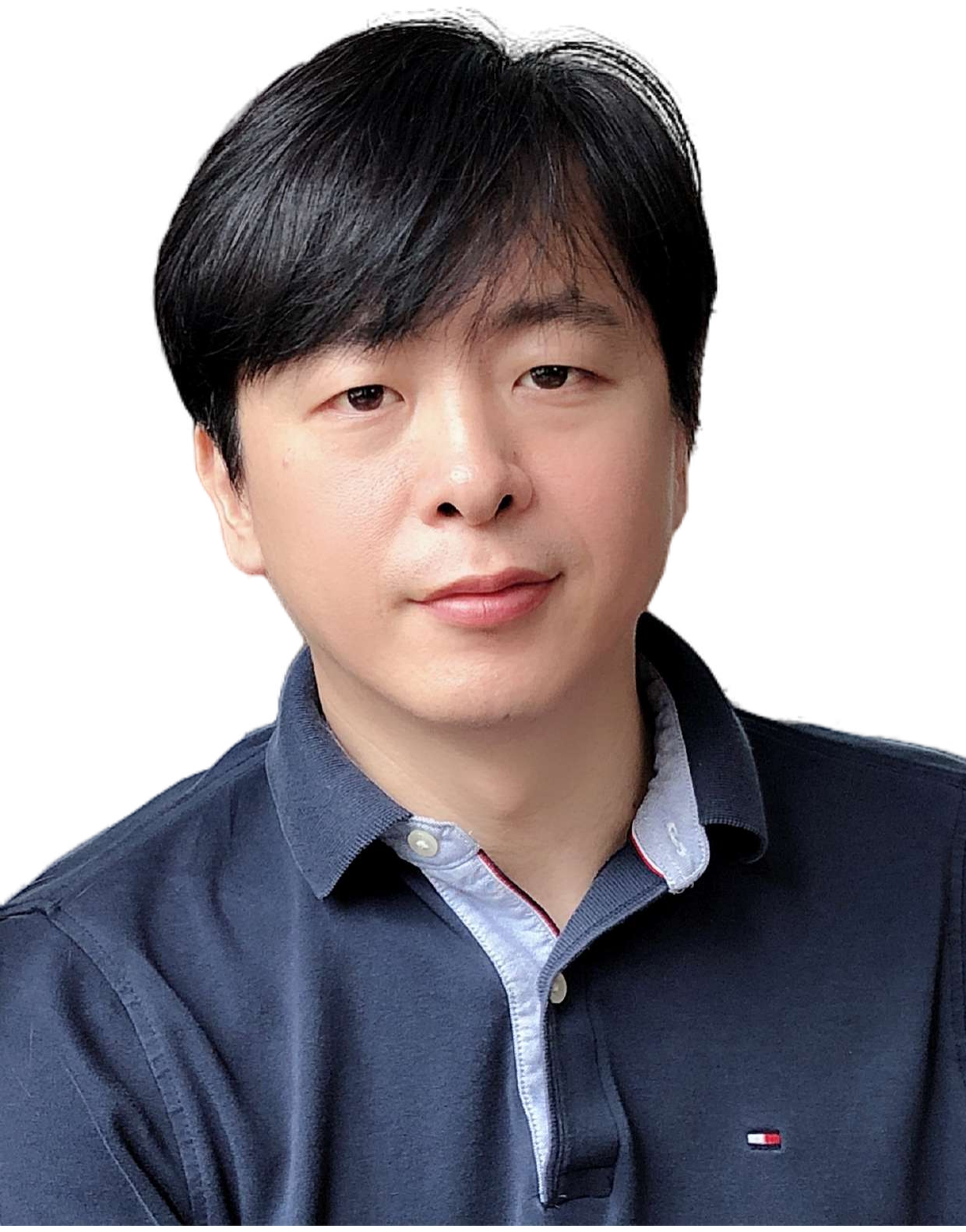}}]
    {Tsung-Yi Ho}
    (F'24)
    is a Professor in the Department of Computer Science and Engineering, The Chinese University of Hong Kong (CUHK). He received his Ph.D. in Electrical Engineering from National Taiwan University in 2005. His research interests include several areas of computing and emerging technologies, especially in the design automation of microfluidic biochips. He was a recipient of the Best Paper Award at the IEEE Transactions on Computer-Aided Design of Integrated Circuits and Systems in 2015. Currently, he serves as the VP Conferences of IEEE CEDA, and the Executive Committee of ASP-DAC and ICCAD. He is a Distinguished Member of ACM and a Fellow of IEEE.
\end{IEEEbiography}